\documentclass[showpacs,amsmath,amssymb,showkeys,pre]{revtex4}
\usepackage{amsmath}          
\usepackage{amssymb}          
\usepackage{epsfig}           

\newcommand{\D}{\mathcal{D}}
\renewcommand{\L}{\mathcal{L}}
\newcommand{\G}{\mathcal{G}}
\newcommand{\F}{\mathcal{F}}
\newcommand{\I}{\mathcal{I}}

\newcommand{\C}{\mathcal{C}}
\newcommand{\R}{\mathbb{R}}
\newcommand{\E}{\mathbb{E}}
\newcommand{\mE}{\mathcal{E}}
\renewcommand{\P}{\mathbb{P}}
\newcommand{\mP}{\mathcal{P}}
\newcommand{\e}{\epsilon}
\newcommand{\s}{\sigma}
\newcommand{\Z}{\mathcal{Z}}
\newcommand{\N}{\mathbb{N}}

\begin{document}

\title{The cavity method for large deviations}

\author{Olivier Rivoire}

\affiliation{Laboratoire de Physique Th\'eorique et Mod\`eles
Statistiques \\
Universit\'e\ Paris-Sud, B\^at. 100, 91405 Orsay Cedex, France}

\date{June, 2005}

\begin{abstract}
A method is introduced for studying large deviations in the context of statistical
physics of disordered systems. The approach, based on an extension
of the cavity method to atypical realizations of the quenched
disorder, allows us to compute exponentially small probabilities (rate functions) over different classes of random graphs. It is illustrated with two
combinatorial optimization problems, the vertex-cover and coloring problems, 
for which the presence of replica symmetry breaking phases is taken into account.
Applications include the analysis of models on adaptive graph
structures.
\end{abstract}

\pacs{79.60.Ht 
, 89.75.Fb. 
}
\keywords{Large deviations, cavity method, optimization.}

\maketitle

\section{Introduction}

The algorithmic complexity of a problem is traditionally measured
on an ensemble of possible inputs (instances) by retaining the
largest time it takes for an algorithm to solve one of the
instances \cite{PapadimitriouSteiglitz82}. Statistical physics
studies have however suggested a different characterization of
hardness, based on the average case rather than the worst case
\cite{MezardParisi87b}. This alternative approach is motivated by
a generic phenomenon of concentration, according to which a
particular instance behaves almost surely as the average case in
the limit of infinite size. This self-averaging property
is common to many disordered systems whose environment is
specified by a quenched random variable from a prescribed
ensemble~: in the thermodynamical limit, the properties of a
sample tend to be independent of the particular realization of the
disorder. Due to this parallel, the methods first developed for
physical disordered systems have been successfully applied to
combinatorial optimization problems \cite{MezardParisi87b}. The interest in 
typical properties is however far from limited to physics or optimization
; information \cite{CoverThomas91} and graph theories
\cite{Bollobas01} are two other major fields where they play a key
role, as illustrated by the seminal works of Shannon
\cite{Shannon48} and Erd\H{o}s-R\'enyi \cite{ErdosRenyi60}
respectively.

In any practical implementation however, optimization or coding
theories face large but finite system sizes. In such situations,
controlling the deviations from the typical case becomes of
primarily practical interest. The scope of large deviations theory
\cite{DemboZeitouni93} is precisely to evaluate the probability of
the rare events associated with such finite size effects. In
addition, and despite the tremendous number of their elements,
large deviations theory has also a direct relevance for physical
systems : as will be explained, it underlies the thermodynamics of
systems whose configuration space is constituted of different
realizations of the quenched disorder. Consequently, when the
disorder consists of an ensemble of random graphs, it allows to
solve models with variable topologies. A large deviations analysis
is thus useful to address the adaptability of constrained systems,
with a physical example being covalent molecular networks subject
to stress \cite{BarreBishop05b,Rivoire06}.

A particularly powerful method for computing typical properties of
disordered systems is the celebrated replica method
\cite{MezardParisi87b}. The cavity method
\cite{MezardParisi87b} provides an alternative tool
that yields equivalent results but has two advantages over the
replica method : it is based on assumptions formulated explicitly,
and it applies to particular instances. In the present paper, we
develop an approach of large deviations based on the cavity
method, which we call the large deviations cavity method (LDCM). 
If large deviations have only recently raised an
interest in the statistical physics community 
\cite{DerridaLebowitz02,MontanariZecchina02,AndreanovBarbieri04,
BarreBouchet04,EngelMonasson04,Rivoire04},
they have a much longer history in probability theory
\cite{DemboZeitouni93}, where they are notably used to rigorously
solve statistical mechanics models \cite{Ellis85}. At
variance with this mathematical tradition, the method exposed here
is non-rigorous and only provides a coherent heuristic framework
for obtaining quantitative predictions, namely computing rate
functions assuming a large deviations principle indeed holds.
Nonetheless, as for the "typical" cavity method of M\'ezard and
Parisi \cite{MezardParisi01,MezardParisi03} that we will recover as a
particular case, the LDCM is hoped to be
amenable to rigorous studies.

The paper is organized as follows. The first section is devoted to
introducing some basic elements of combinatorial optimization and
large deviations theories, with an emphasis on their links with
statistical mechanics. The second section presents the LDCM
 in its simplest, "replica symmetric",
form : we start by rederiving in a cavity-like fashion Cram\'er's
theorem, the most elementary result in large deviations theory, and then
discuss different graph ensembles, with explicit calculations on
the vertex-cover problem. The third section deals with systems
having a non trivial internal structure : we notably generalize
the large deviations approach to systems displaying a glassy
"replica symmetry breaking" phase, a situation that we illustrate
in details with the coloring problem. A conclusion closes the
paper by suggesting some possible applications.

\section{Optimization problems and large deviations}

\subsection{Optimization problems}\label{sec:opt}

The field of combinatorial optimization provides a broad class of
disordered systems, which we use here to illustrate the
potentialities of the LDCM. We therefore
start with a brief introduction to this subject (see e.g.
\cite{PapadimitriouSteiglitz82} for more details). Combinatorial
optimization is primarily concerned with minimizing cost
functions, $\mE:\C\mapsto \R$, over some discrete configuration
space $\C$. In view of quantifying their algorithmic complexity,
optimization problems are defined over an ensemble $\I$ of
instances $I$, each associated with a cost function $\mE_I$ . In
particular, many combinatorial problems are defined over ensemble
of graphs \cite{GareyJohnson79}, in which case an instance $I$ is
specified by a graph $G$, that is a set of $N$ nodes
$i=1,\dots,N$ associated with a subset of the pairs
$\{(i,j)\}_{i\neq j}$, defining its edges.

Two prototypical examples will serve as illustration. The first
one is the vertex-cover problem \cite{GareyJohnson79}, also known
as independent set, which consists, given a graph $G$, in blackening
as many of its nodes as possible while never blackening two
connected nodes. The second, quite similar, example is the
coloring problem \cite{GareyJohnson79} which asks, given a graph
$G$ and $q$ colors, whether it is possible to assign a color to
each node of $G$ so that no two adjacent nodes have same color.
The coloring problem is strictly speaking a decision problem (the
answer must be yes or no) but it is directly related to the
optimization problem of minimizing the number of edges having two
end-nodes sharing a same color : if the minimum is zero, the graph
is colorable, otherwise it is uncolorable.

From the statistical physics viewpoint, the cost $\mE_I[\sigma]$
represents the energy of a configuration $\sigma\in\C$, and the
minimal cost $E_I=\min_\sigma \mE_I[\sigma]$ corresponds to the
ground-state energy of the disordered system having quenched
disorder $I$. In this context, it is usual to introduce an inverse
temperature $\beta$ and a free energy density $f_I(\beta)$ defined
by $f_I(\beta)=-\ln[\sum_{\sigma}\exp(-\beta
\mE_I[\sigma])]/(\beta N)$, such that the ground-state energy
density $\e_I=E_I/N$ is given by the $\beta\to\infty$ limit,
$\e_I=\lim_{\beta\to\infty}f_I(\beta)$. For the coloring problem,
the associated finite temperature system is known as the
antiferromagnetic Potts model \cite{Wu82}, while it is called the
hard-core model \cite{WeigtHartmann01} for vertex-cover (with
$\beta$ representing a chemical potential).

For these two examples, the configuration space $\C$ is discrete :
$\C$ can be taken as $\{0,1\}^N$ for vertex-cover, with $\s_i=0$
and $\s_i=1$ corresponding respectively to uncovered (white) and
covered (black) nodes, and as $\{1,\dots,q\}^N$ for coloring with
$\s_i\in\{1,\dots,q\}$ now representing the color assigned to $i$.
Different ensembles $\G$ of graphs, defining different sets of
instances $\I$, can be introduced. The cavity approach followed
here applies to any ensemble of locally "tree-like" graphs, that is graphs
whose degrees (the number of nodes to which a given node is
connected) remain finite when $N\to\infty$. Three random graphs ensembles
will be specifically addressed here. The first one,
noted $\tilde{\G}^{(\gamma)}_N$, is the set of graphs with $N$
nodes where each edge as a probability $\gamma/N$ to be present,
and is known as the binomial model or Erd\H{o}s-R\'enyi ensemble
\cite{Bollobas01}. The second one, noted $\bar{\G}^{(\gamma)}_N$,
and called the uniform model \cite{Bollobas01}, is the set of
graphs with $N$ nodes and $M=\gamma N/2$ edges. Finally, the third
one is defined through the degree distribution $p(k)$ of the nodes
of its graphs \cite{NewmanStrogatz01} : each of the $N$ nodes has
degree $k$ with independent probability $p(k)$ and the edges are
drawn at random subject to that constraint. This last class notably
includes random regular graphs \cite{Bollobas01}, for which
$p(k)=\delta_{r,k}$, and power-law distributed graphs, for which
$p(k)\sim k^\tau$ (with an appropriate cut-off to insure
normalization). Here we will also consider the Poissonian model
noted $\hat{\G}^{(\gamma)}_N$ and defined by
$p(k)=\gamma^ke^{-\gamma}/k!$. In the $N\to\infty$ limit, the
binomial and uniform models $\tilde{\G}^{(\gamma)}_N$ and
$\bar{\G}^{(\gamma)}_N$ share with $\hat{\G}^{(\gamma)}_N$ the
same Poisson degree distribution. This equivalence between the
three models extends to the typical properties of optimization
problems defined on them but, as will be shown, does not hold
for atypical features.

From the point of view of computational complexity, an important
parameter is the size $N$ of the instances, which, in the case of
diluted graphs, is taken as the number of nodes. As seen on the
vertex-cover and coloring problems, the size of the configuration
space $\C$ over which optimization is to be performed increases
exponentially with $N$, precluding any na\"{\i}ve exhaustive
search for large $N$ and possibly making the problem highly
non-trivial. In fact, both the vertex-cover and coloring problems
are known to be NP-hard in the worst case, implying that no
algorithm is known that can solve all instances of these problems
in a time growing polynomially with $N$
\cite{PapadimitriouSteiglitz82}.

As stressed in the introduction, the focus on typical instances
advocated by statistical physics is justified by the
self-averaging property : when it holds for an ensemble $\I$ of
instances, there is a typical value of the ground-state energy
density $\bar{\e}$ such that, for any $\varepsilon>0$, the
probability $\mathbb{P}[|E_I/N-\bar{\e}|>\varepsilon]$ for the
optimum $E_I$ to deviate from $N\bar{\e}$ goes to zero,
$\mathbb{P}[|E_I/N-\bar{\e}|>\varepsilon]\to 0$ as $N\to\infty$.
Informally, large problems then typically all share a common
optimum, which, physically, can often be traced back to the
equivalence of their local properties. The "typical" cavity method
\cite{MezardParisi01, MezardParisi03} have been developed
precisely to compute the most probable value $\bar{\e}$ for
problems on random graphs. The LDCM presented
here is an extension of this approach that allows to evaluate the
$N$ and $\varepsilon$ dependencies of vanishing probabilities such
as $\mathbb{P}[|E_I/N-\bar{\e}|>\varepsilon]$.

\subsection{Large deviations}

For finite $N$, an instance has always a finite probability to
deviate from the typical case. The so-called large deviations
\cite{DemboZeitouni93} refer to the extensive deviations from
$N\bar{\e}$, of order $O(N)$, as distinguished from the small,
subextensive deviations from $N\bar{\e}$, of order $o(N)$, like
for example $O(\sqrt{N})$ fluctuations (see however below for a
relation between the two). The present method is based on an
Ansatz, according to which large deviations are exponentially
small in the size $N$ of the instances, that is, the probability
$\P_N[E_I]$ for an instance $I$ taken out of the ensemble $\I$ to
have an optimal cost $E_I$ is supposed to satisfy
\begin{equation}\label{eq:ansatz}
\P_N[E_I=N \e]\asymp e^{-N L(\e)},
\end{equation}
where the symbol $\asymp$ stands here and in the sequel for an
exponential equivalence defined as $\lim_{N\to\infty}\ln(\P[E_I=N
\e])/N = -L(\e)$. $L(\e)$ is called a rate function, or large
deviations function, and, in the simplest cases, is strictly
positive, except for the typical value $\bar{\e}$ where it
achieves its zero minimum. The Ansatz (\ref{eq:ansatz}) is
known to indeed hold in the solvable case where
$\mE_I$ consists of a sum of independent identically distributed
variables (Cram\'er's theorem, see Sec.~\ref{sec:cramer}), and this result
is robust to the presence of weak correlations among the
variables (G\"atner-Ellis theorem, to be stated below) \cite{DemboZeitouni93}. 
More precisely, the relation~(\ref{eq:ansatz}) corresponds in the mathematical
literature to the "large deviations principle"
\cite{DemboZeitouni93}, which, in its simplest form, can be stated
as follows:

{\bf Large deviations principle:} The sequence $\{\e_N\}_{N\in\N}$
of real valued random variables is said to satisfy the large
deviations principle, with rate function $L:\R\to\R^+\cup\{\infty\}$, if\\
$(i)$ $\forall M\geq 0,\quad \{x\in\R : L(x)\leq M\} {\rm \ is\ compact}$,\\
$(ii)$ for all closed subset $F$ of $\R$, and all open subset $O$ of $\R$,
\begin{equation}
\limsup_{N\to\infty}\frac{1}{N}\P_N[\e_N\in F]\leq -L(F),\quad\quad
\liminf_{N\to\infty}\frac{1}{N}\P_N[\e_N\in O]\geq -L(O).
\end{equation}

We point out right away that counterexamples are easily found for
which the previous Ansatz does not hold. In the field of
spin-glass models, they include the two most celebrated models,
the random energy model \cite{Derrida80} and the SK model
\cite{SherringtonKirkpatrick75}. For the random energy model,
elementary calculations \cite{AndreanovBarbieri04, Rivoire04}
indeed give
\begin{equation} \P_N(\e)\asymp
\begin{cases}
e^{-e^{Ns(\e)}} & \text{if $\e>\bar{\e}$}, \\
e^{Ns(\e)} & \text{if $\e<\bar{\e}$},
\end{cases}
\end{equation}
with $s(\e)=\ln 2-\e^2$ and $\bar{\e}=-\sqrt{\ln 2}$. For the SK
model, numerical studies \cite{AndreanovBarbieri04} also suggest
different scalings on both sides of the typical value, that is
$\P_N(\e)\asymp \exp[-N^aL(\e)]$ with $a\simeq 1.2$ when
$\e<\bar{\e}$, but $a\simeq 1.5$ when $\e>\bar{\e}$. However, in a
variety of other spin-glass models, notably including models on
diluted random graphs, the Ansatz (\ref{eq:ansatz}) is supported
by numerical evidence \cite{AndreanovBarbieri04}.

From the analytical viewpoint, rate functions in the context of
optimization problems have been studied by Montanari
\cite{MontanariZecchina02,Montanari02}, using the replica method.
The results he obtained for the vertex-cover problem
\cite{MontanariZecchina02,Montanari02} are strictly equivalent to
the ones to be derived here from the cavity method. Yet, as for the
typical case, the cavity approach has the advantages over the
replica method to offer a more transparent derivation, and to open
the way to algorithmic implementations on particular systems. For
a model at finite temperature $1/\beta$, the replica method
basically consists in inferring rate functions from the knowledge
of the moments $\mathbb{E}[Z^n_I]$ of the partition function
$Z_I(\beta)=\sum_\s\exp(-\beta \mE_I[\s])$, with $\mathbb{E}[\cdot]$
referring to the average over the disorder, that is the different
instances $I$. As far as no replica symmetry breaking is involved,
this procedure is motivated by the following rigorous result
\cite{DemboZeitouni93} :

{\bf G\"artner-Ellis theorem :} Let $\{\e_N\}_{N\in\N}$ be a
sequence of real valued random variables and let $\F:\R\to\R$ be
defined by
\begin{equation}\label{eq:GE}
\F(y)=-\lim_{N\to\infty}\frac{1}{N}\ln \E[e^{-y N\e_N}].
\end{equation}
If $\F(y)$ exists, is finite and differentiable for every
$y\in\R$, then the sequence $\{\e_N\}_{N\in\N}$ satisfies the
large deviations principle with rate function $L(\e)$ given by the
Legendre transform of $\F(y)$,
\begin{equation}
-L(\e)=\inf_{y\in\R}[y\e-\F(y)],
\end{equation}
where minus signs are introduced to match usual conventions in
statistical physics. We stress that for sake of simplicity, this theorem is stated here
with much stronger hypothesis than necessary ; in particular the
assumptions about the finiteness and differentiability of $\F$
can be relaxed \cite{DemboZeitouni93}.

To apply the replica method to optimization problems, the limit
$\beta\to\infty$ has to be considered, and, to obtain non trivial
results in this limit, the replica number $n$ must be rescaled
with $\beta$, such that $\beta\to\infty$ and $n\to 0$ with
$y=n/\beta$ finite. In this limit, the replica potential $\F(y)$
coincides with the function introduced in G\"artner-Ellis theorem, Eq.~(\ref{eq:GE}),
\begin{equation}
e^{-N\F(y)}=\lim_{\beta\to\infty}\mathbb{E}[Z^{y/\beta}]=\mathbb{E}[e^{-y
E}] =\int e^{-N[L(\e)+y\e]}d\e.
\end{equation}
While proceeding differently, the cavity method to be presented
will lead to the same rate function, again specified as the
Legendre transform of the potential $\F(y)$. Although both the
replica and the cavity methods, based on Legendre transformations,
naturally yield convex functions, it should be stressed that
convexity is not a necessary feature of rate functions. In fact,
non-convex rate functions are associated with phase transitions
and are therefore encountered in many models of interest from the
statistical mechanics point of view \cite{Ellis85}.

Large deviations deal with exponentially small probability and may
appear as only an extreme feature of finite size effects, while a
more refined description would consist in the complete probability
distribution of $E_I$ over $\I$. Interestingly, small fluctuations
can be extracted from the knowledge of the rate function
near its typical minimum. More precisely, the potential $\F(y)$
yields the cumulants of $\langle(E_N)^k\rangle_c$,
\begin{equation}
\langle(E_N)^k\rangle_c=-N\frac{\partial^k\F}{\partial y^k}(y=0),
\end{equation}
where, as usual, the cumulants $\langle X^k\rangle_c$ of a random
variable $X$ are defined by
$\ln\mathbb{E}[e^{tX}]=\sum_{k=1}^\infty \frac{t^k}{k!}\langle
X^k\rangle_c$. In particular, the Ansatz (\ref{eq:ansatz})
predicts the variance of the small fluctuations to be generically
of order $\sqrt{N}$, as given by the central limit theorem in the
case of a sum of independent identically distributed variables.

\subsection{Statistical mechanics interpretation}\label{sec:interpret}

On top of their own mathematical interest, large deviations are of
direct relevance to statistical mechanics studies. In the context
of optimization problems, rate functions can indeed be interpreted
as defining an entropy on the space of the instances $\I$,
corresponding to a thermodynamics over the quenched disorder. This
relation, formalized by Sanov's theorem \cite{Ellis85}, is
presented here in the restricted context where $\I$ is a class of
graphs associated with a given optimization problem.

Viewing the ensemble of random graphs $\mathcal{G}_N$ as a phase
space, each graph $G\in\mathcal{G}_N$ defines a configuration to
which is associated the ground-state energy $E_G$, that is, the
optimal cost for the optimization problem on $G$. If
$|\mathcal{G}_N|\equiv e^{Ns_0}$ denotes the cardinality of
$\mathcal{G}_N$, the microcanonical entropy $s(\e)$ of the system
is given by
\begin{equation}
e^{Ns(\e)}= \#\{G\in \mathcal{G}_N ; E_G=N\e\}=
\#\mathcal{G}_N\times\mathbb{P}_N[E_G=N\e]\asymp e^{N[s_0-L(\e)]},
\end{equation}
where $\# A$ denotes the cardinality of the set $A$.
Thus, up to a linear transformation, the rate function
$L(\e)$ is nothing but the microcanonical entropy $s(\e)$,
\begin{equation}\label{eq:s}
s(\e)=s_0-L(\e).
\end{equation}
Within this picture, the parameter $y$ appearing in the replica method
and G\"artner-Ellis theorem represents the external inverse 
temperature that allows to study statistical mechanics on the
configuration space spanned by the graphs, $y\equiv-\partial_\e
L(\e)=\partial_\e s(\e)$ ($y$ must be distinguished from the internal inverse
temperature $\beta$ which is set to infinity in the context of optimization).
By construction, this space has no more
quenched disorder, and a large deviations analysis appears as the
statistical mechanics analysis of a pure system at finite inverse
temperature $y$. From the opposite viewpoint, large deviations
theory thus provides a meaning for negative temperatures, $y<0$.
Finally, the typical case is given by the infinite temperature
limit, $y=0$, as prescribed by replica theory.

We stress however that the possibility of deriving the
thermodynamics of the system at inverse temperature $y$ from the
knowledge of its microcanonical entropy $s(\e)$ is based on the
equivalence between the microcanonical and canonical ensembles in the
thermodynamical limit, which can not always be taken for
granted. In presence of non-convex rate functions indeed, the two
ensembles become inequivalent, and a first order transition
occurs, whose description requires a Maxwell construction ; such a
construction in the context of large deviations for random graph
has been recently described in \cite{EngelMonasson04}.

We have restricted so far to the simplest case where the measure
over the quenched disorder is an uniform measure over an ensemble
of graphs, but more complicated structures can be considered as
well. In particular, the disorder can have different origins, as
with spin-glass models \cite{MezardParisi87b} or $K$-SAT
optimization problems \cite{PapadimitriouSteiglitz82}, where in addition
to the graph structure, the quenched disorder comprises the
specification of some random couplings between the variables. In
this case, an instance $I$ of the problem is first defined by
selecting a graph $G$ and then by choosing the couplings $J$.
Large deviations can be taken with respect to $J$ at
fixed $G$ : for typical graphs $\bar{G}$, the effective system
still contains a quenched disorder (the graph) which can be
handled with the usual techniques of disordered systems, but if
atypical graphs have to be addressed as well, a second
temperature needs to be introduced. The two temperatures are in
such a case associated with two levels of probability
distributions, in a construction formally identical to Parisi's
hierarchical scheme for handling replica symmetry breaking, as will be
discussed in Sec.~\ref{sec:2step}.
The same scheme also applies when going to lower levels to
describe the internal structure of a given instance. This will be
exemplified in Sec.~\ref{sec:finiteT} where we discuss the
implications of working with a finite temperature on the
instances, or working with optimization problems displaying a replica
symmetry breaking phase.

\section{The large deviations cavity method }

The "typical" cavity method, as developed by M\'ezard and Parisi
\cite{MezardParisi01,MezardParisi03}, applies to a given instance
$I$ and addresses the structure of its phase space, that is the
organization of the configurations $\s\in\C$ as a function of
their energy density $\mE[\sigma]/N$. The method can handle either
a structure composed of a unique set (or a finite number of sets)
of connected solutions, called a {\it replica symmetric} (RS)
phase, or a structure composed of many disconnected clusters of
configurations, called a {\it one-step replica symmetry breaking}
(1RSB) phase \cite{MezardParisi87b}. In the latter case, the
crucial assumption is made that the number of clusters with a
given energy density $\e$ is exponential in $N$,
\begin{equation}\label{eq:ansatz1rsb}
\mathcal{N}_{\rm clusters}(\e)\asymp e^{N\Sigma(\e)}.
\end{equation}
The 1RSB cavity method is specifically designed to compute the
function $\Sigma(\e)$, called the {\it complexity}, with the
particular RS case corresponding to $\Sigma=0$ \cite{Monasson95}. 
The formal analogy
between the 1RSB Ansatz (\ref{eq:ansatz1rsb}) defining the
complexity $\Sigma(\e)$ and the large deviations Ansatz
(\ref{eq:ansatz}) defining the rate function $L(\e)$ is at the root of 
the possibility to extend the typical cavity method yielding $\Sigma(\e)$ 
to an atypical version
yielding $L(\e)$. To emphasize further the parallel, we introduce
the function $\L(\e)$ defined as $\L(\e)\equiv -L(\e)$, such that
$\L(\e)$ plays in the LDCM a role
formally identical to the complexity in the typical cavity method:
\begin{equation}\label{eq:reAnsatz}
\P_N(E=N\e)\asymp e^{N\L(\e)}.
\end{equation}

The analogy between the complexity $\Sigma(\e)$ and the rate
function $L(\e)$ should not be taken for a coincidence: the
complexity is fundamentally nothing but a rate function [or more
accurately the entropy associated to it, as in Aq.~(\ref{eq:s})],
which describes the large deviations of the energy over the different
clusters of solutions. From this point of view, further elaborated
in \cite{these}, the 1RSB cavity method is itself a
large deviations method, acting on the self-generated (glassy)
"internal disorder" of a given sample. For glassy optimization
problems, being able to address such large deviations is crucial
since ground-state clusters are atypical, that is, exponentially
less numerous than clusters with higher energies. These atypical
ground-state clusters must be obtained by correctly tuning the
"internal inverse temperature", noted $\mu$ in this context.
Remarkably, while the LDCM to be
presented will also apply to the typical case $y=0$, the 1RSB
cavity method is in general not able to describe the complete
complexity curve $\Sigma(\e)$, and notably fails to
describe the most numerous, typical clusters, corresponding to
$\mu=0$ \cite{footnoteSUSY}.

Our presentation of the LDCM will
follow closely the presentation of M\'ezard and Parisi of their
typical cavity method \cite{MezardParisi01,MezardParisi03}, but
major differences will show up in the way averages over the
disorder are performed. To start with, we consider the simplest
case where the underlying optimization problem is assumed to be
itself RS i.e., with no clustering induced by its internal
disorder.

\subsection{The cavity approach to Cram\'er's theorem}\label{sec:cramer}

Although its most interesting applications involve random graphs,
the cavity method is not restricted to this particular geometry.
As an illustration of the ideas in their simplest setting, we
consider the case, with no geometry, of a system made of $N$
independent elements, each contributing to the total energy
$\mE_N$ by a random amount $X_i$. In other words, we consider here
large deviations in the sum of independent identically distributed
random variables. For such a system, the typical energy density
follows from the law of large numbers, which, assuming the
distribution $\rho(X)$ of the $X_i$'s to have a finite first
moment, is $\bar{\e}=\mathbb{E}[X]\equiv\int x\rho(x)dx$. Large
deviations are concerned with deviations from the prediction
$\e_N/N= \bar{\e}$ and, for a sum of independent variables, are
completely specified by Cram\'er's theorem, both a generalization of
the law of large numbers and a corollary of G\"artner-Ellis
theorem \cite{DemboZeitouni93}.

{\bf Cram\'er's theorem :} Let the sequence $\{\e_N\}_{N\in\N}$ of
real random variables be given by $\e_N=(\sum_{i=1}^N X_i)/N$
where the $\{X_i\}_i$ are independently identically distributed
real random variables. If $\E[e^{-y
X}]$ is finite for all $y\in\R$, then $\{\e_N\}_{N\in\N}$
satisfies the large deviations principle with rate function
$L:\R\to\R$ defined as Legendre transform of $\F:\R\to\R$ given by
$\F(y)\equiv-\ln\E[e^{-yX}]$, that is
\begin{equation}
-L(\e)\equiv\L(\e)=\inf_{y\in\R}[y\e-\F(y)].
\end{equation}

The basic idea behind the cavity approach is to estimate the
change of the system upon addition of a new variable (or,
equivalently, upon removal of a variable, hence the name
"cavity"). Let $\mE_N$ be the extensive energy,
$\mE_N=\sum_{i=1}^N X_i$. By virtue of the assumed independence of
the $X_i$, the probability distribution for
$\mE_{N+1}=\mE_N+X_{N+1}$ is given by a convolution of those of
$\mE_N$ and $X_{N+1}$, which, with the Ansatz (\ref{eq:reAnsatz}),
reads
\begin{equation}\label{eq:cavCramer}
\mathbb{P}_{N+1}(\mE_{N+1}=E)=e^{(N+1)\mathcal{L}\left(\frac{E}{N+1}\right)}
=\mathbb{E}_X[\mathbb{P}_N(\mE_N=E-X)]=\int\rho(\Delta
E)e^{N\mathcal{L}\left(\frac{E-\Delta E}{N}\right)}d\Delta E.
\end{equation}
Assuming a smooth behavior of $\mathcal{L}$, we write for large
$N$,
\begin{equation}
\begin{split}
&(N+1)\mathcal{L}\left(\frac{E}{N+1}\right)=
N\mathcal{L}(\e)+\mathcal{L}(\e)-\partial_\e\mathcal{L}(\e)+O(1),\\
&N\mathcal{L}\left(\frac{E-\Delta E}{N}\right)=
N\mathcal{L}(\e)-\Delta E\partial_\e\mathcal{L}(\e)+O(1),
\end{split}
\end{equation}
where $\e\equiv E/N$. Setting $y\equiv
\partial_\e\mathcal{L}(\e)$ thus yields
\begin{equation}
\F(y)\equiv y\e-\mathcal{L}(\e)=-\ln\mathbb{E}[e^{-yX}].
\end{equation}
We conclude that $\mathcal{L}(\e)$ is given by the Legendre
transform of the potential $\F(y)$,
\begin{equation}
\begin{split}
&\mathcal{L}(\e)=\e y-\F(y),\\
& \e=\partial_y\F(y),
\end{split}
\end{equation}
as prescribed by Cram\'er's theorem.

\subsection{Poissonian random graphs}

We consider now models defined on random graphs, first under the
assumption that the internal structure of an instance is replica
symmetric (RS). As a further simplification (to be relaxed later
on, as for the RS hypothesis), we assume that the only source of
quenched disorder lies in the graph structure, as it is the case
for the vertex-cover and coloring problems. We consider here
simultaneously the three ensembles of random graphs,
$\tilde{\G}^{(\gamma)}_N$, $\bar{\G}^{(\gamma)}_N$ and
$\hat{\G}^{(\gamma)}_N$ defined in \ref{sec:opt} and hereafter
generically referred to as $\G_N^{(\gamma)}$. In the $N\to\infty$
limit, the degrees of graphs in $\G_N^{(\gamma)}$ have same
limiting distribution $\pi_\gamma$, where $\pi_\gamma$ denotes the
Poisson distribution with mean $\gamma$,
$\pi_\gamma(k)=\gamma^ke^{-\gamma}/k!$.

Taking a graph uniformly at random in $\G_N^{(\gamma)}$ defines
the measure over the quenched disorder with respect to which large
deviations are evaluated. Following the basic principles of the
cavity method, we consider the changes in the system when its size
is increased from $N$ to $N+1$. The first idea would be to
construct uniformly at random a graph $G_{N+1}$ in
$\G_{N+1}^{(\gamma)}$ from a graph $G_N$ randomly chosen in
$\G_N^{(\gamma)}$, by connecting a new node to $k$ nodes of $G_N$,
with $k$ taken with the distribution $\pi_\gamma$. This
construction is however too na\"\i ve, since, if the initial graph
was in $\bar{\G}^{(\gamma)}_N$ for example, its extension is in
$\bar{\G}^{(\gamma')}_{N+1}$, with
$\gamma'/2=(M+k)/(N+1)\neq\gamma/2$, where $M=\gamma N/2$ denotes the number of edges
in $G_N$. However, for the three
models, it appears that this construction yields a graph uniformly
at random in $\G_{N+1}^{(\gamma')}$, with
$\gamma'=\gamma+\chi(\gamma,k)/N$, where we have obtained for
$\bar{\G}^{(\gamma)}_N$ that $\bar{\chi}(\gamma,k)=2k-\gamma$. For $\tilde{\G}^{(\gamma)}_N$,
after addition of the new node the probability for an edge to be
present is still $\gamma/N$ and not $\gamma/(N+1)$, so that it is
described by $\gamma'$ satisfying $\gamma'/(N+1)=\gamma/N$,
yielding $\tilde{\chi}(\gamma,k)=\gamma$. Finally for
$\hat{\G}^{(\gamma)}_N$, after addition of the new site, the
distribution of the degrees $\pi_\gamma(K)$ is modified to
$(1-k/N)\pi_\gamma(K)+(k/N)\pi_\gamma(K-1)$ since $k$ of the nodes
of $G_N$ receive an additional edge ; this leads to an effective
distribution $\pi_{\gamma'}(K)$ with $\gamma'=\gamma+k/N$, so that
$\hat{\chi}(k,\gamma)=k$. To sum up, we obtained
\begin{eqnarray*}\label{eq:chi}
\bar{\chi}(\gamma,k)=&2k-\gamma &\text{(Uniform model }\bar{\G}^{(\gamma)}_N),\\
\tilde{\chi}(\gamma,k)=&\gamma &\text{(Binomial model }\tilde{\G}^{(\gamma)}_N),\\
\hat{\chi}(\gamma,k)=& k &\text{(Poissonian model }\hat{\G}^{(\gamma)}_N).\\
\end{eqnarray*}
The fact that in all cases
$\langle\chi(\gamma,k)\rangle=\gamma$, with the average
$\langle\cdot\rangle$ taken with respect to $\pi_\gamma$,
reflects the equivalence of the typical properties between the
three models.

When a node is added, the ground-state energy is shifted by an
amount $\Delta E$. Conditioned to the fact that the new node is
connected to $k$ other nodes, this shift has distribution
$P_n^{(k)}(\Delta E)$ from graph to graph (and from node to node
on a given graph). Given $P_n^{(k)}(\Delta E)$, the argument followed in
Eq.~(\ref{eq:cavCramer}) can essentially be repeated,
\begin{equation}\label{eq:argument}
\begin{split}
&e^{(N+1)\L(E/(N+1),\gamma)}\asymp e^{N\L
(\e)}e^{\L(\e,\gamma)-y\e}=e^{N\L
(\e)}e^{-\F(y,\gamma)}\\
&=\sum_{k=0}^\infty \pi_\gamma(k) \int d\Delta E P_n^{(k)}(\Delta
E) e^{N\L[(E-\Delta E)/N,\gamma-\chi(\gamma,k)/N]}\asymp
e^{N\L (\e)}\sum_{k=0}^\infty \pi_\gamma(k) \int d\Delta E
P_n^{(k)}(\Delta E) e^{-y\Delta E}e^{-z\chi(\gamma,k)},
\end{split}
\end{equation}
with the notations $\e\equiv E/N$,
$y\equiv\partial_\e\L(\e,\gamma)$, $z\equiv
\partial_\gamma \L(\e,\gamma)$ and
$\F(y,\gamma)=y\e-\L(\e)$. Eq.~(\ref{eq:argument}) gives the Legendre transform of the rate function, $\F(y,\gamma)$, as a function of $y$ and $z$.
To determine $z$, we need consider the energy shift $\Delta E$ due
to a link addition, having distribution $P_\ell(\Delta E)$. More
precisely, the average value of the energy shift when
$\gamma\to\gamma+1/N$ at fixed number of nodes $N$ is required,
which is obtained by adding $k$ new edges with an appropriate
distribution $\sigma(k)$. For $\bar{\G}^{(\gamma)}_N$, adding a
single edge results in $\gamma'/2=(M+1)/N=\gamma/2+1/N$ so we take
formally $\sigma=\delta_{1/2}$, where
$\delta_\theta(k)=\delta_{k,\theta}$ denotes the Kronecker symbol (this
non-integer prescription could be avoided as in
\cite{MezardParisi03} by adding two nodes at once instead of
one). For $\tilde{\G}^{(\gamma)}_N$, we take $\sigma=\pi_{1/2}$
because it corresponds to the distribution of the number of added
edges when each of the $\sim N^2/2$ edges has a probability
$1/N^2$ to be present in the $\tilde{\G}^{(\gamma+1/N)}_N$ graph,
but absent in the $\tilde{\G}^{(\gamma)}_N$ one. Finally for
$\hat{\G}^{(\gamma)}_N$, the addition of one edge leads to
$\gamma'=\gamma+2/N$ so that formally $\sigma=\delta_{1/2}$ as in
the uniform model. A $1/N$ expansion of $\exp
[N\L(\e,\gamma+1/N)]$ then yields $z=\partial_\gamma
\L(\e,\gamma)$ as
\begin{equation}\label{eq:z}
e^z=\sum_{k=0}^\infty \sigma(k)\left(\int d\Delta E P_\ell(\Delta
E)e^{-y\Delta E} \right)^k,
\end{equation}
with as derived just above,
\begin{eqnarray*}\label{eq:sigma}
\bar{\sigma}(k)=&\delta_{1/2}(k) &\text{(Uniform model }\bar{\G}^{(\gamma)}_N),\\
\tilde{\sigma}(k)=&\pi_{1/2}(k) &\text{(Binomial model }\tilde{\G}^{(\gamma)}_N),\\
\hat{\sigma}(k)=&\delta_{1/2}(k) &\text{(Poissonian model }\hat{\G}^{(\gamma)}_N).\\
\end{eqnarray*}

\begin{figure}
\centering \epsfig{file=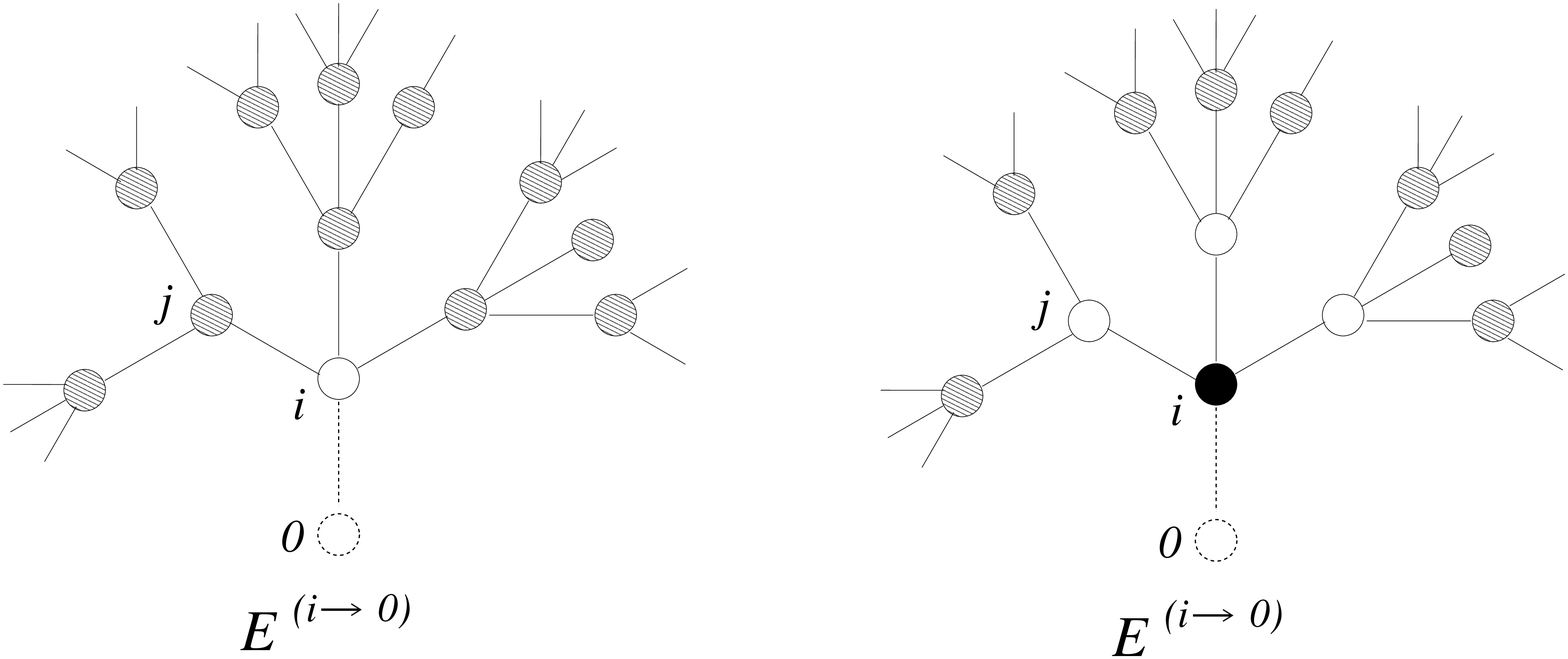,width=10cm}
\caption{Rooted trees in the vertex-cover problem. This
figure illustrates the recursion expressed by
Eq.~(\ref{eq:basicvc}) : on the left tree the root $i$ (in absence
of $0$) is constraint to be uncovered, while on the right tree it
is constraint to be covered. The coloring in gray indicates that
the nodes are neither constraint to be uncovered, nor to be
covered.}\label{fig:rtree}
\end{figure}

As in the typical cavity method
\cite{MezardParisi01,MezardParisi03}, the distributions
$P_n^{(k)}(\Delta E)$ and $P_\ell(\Delta E)$ can be calculated by
means of cavity fields. The fundamental assumption made at this
stage is that the nodes to which a new node is added are
independent in the absence of the added node. Under this
assumption, the problem on a random graph is reduced to a problem
on a tree with self-consistent boundary conditions (the so-called
Bethe lattice). While the same procedure applies to other
optimization problems, we restrict here for simplicity to
the vertex-cover problem, for which we take the
ground-state energy as the minimum of non-covered
nodes under the constraint that neighboring nodes cannot be both
covered. A recursion is written for rooted-trees with same degree
distribution $\pi_\gamma(k)$ as the graphs (see Fig.~\ref{fig:rtree}).
In general if the degree distribution is $p(k)$,
the root must be assigned the distribution
$q(k)=(k+1)p(k+1)/\langle k\rangle$, which corresponds to the
probability, when the edge $(i\to 0)$ is chosen at random, that
$i$ has $k$ neighbors in addition to $0$ ; the Poisson
distribution has however the specificity that
$q(k)=p(k)=\pi_\gamma(k)$. For the rooted tree with root-node $i$,
let $E_0^{(i\to 0)}$ be the minimal energy with the root
constrained to be non-covered (white), and $E_1^{(i\to 0)}$ the
minimal energy with the root constrained to be covered (black).
These quantities are related to those associated with the $k$
rooted trees generated by deletion of the edges originating from
$i$ (see Fig.~\ref{fig:rtree}) by
\begin{equation}\label{eq:basicvc}
\begin{split}
&E_0^{(i\to 0)}=1+\sum_{j=1}^k\min(E_0^{(j\to i)},E_1^{(j\to i)}),\\
&E_1^{(i\to 0)}=\sum_{j=1}^k E_0^{(j\to i)}.
\end{split}
\end{equation}
A cavity field is defined for each oriented edge as $h^{(i\to
0)}\equiv E_0^{(i\to 0)}-\min(E_0^{(i\to 0)},E_1^{(i\to 0)})$ ; it
satisfies the relation
\begin{equation}\label{eq:fieldsdef}
h^{(i\to 0)}=\hat{h}^{(k)}(\{h^{(j\to
i)}\})=\max\left(0,1-\sum_{j=1}^kh^{(j\to i)}\right).
\end{equation}
The addition of the new node $i$ is associated with an energy
shift given by
\begin{equation}\label{eq:Enode}
\begin{split}
\Delta E_{\rm node}&=\min(E_0^{(i\to 0)}, E_1^{(i\to
0)})-\sum_{j=1}^k\min(E_0^{(j\to i)},
E_1^{(j\to i)})\\
&=\Delta\hat{E}^{(k)}_n(\{h^{(j\to i)}\})=\min\left(1,\sum_{j=1}^kh^{(j\to i)}\right).
\end{split}
\end{equation}
We will also need the energy shift corresponding to an edge
addition, which reads
\begin{equation}\label{eq:Eedge}
\begin{split}
\Delta E_{\rm edge}&=\min(E_0^{(1\to 2)}+E_0^{(2\to 1)},
E_0^{(1\to 2)}+E_1^{(2\to 1)},E_1^{(1\to 2)}+E_0^{(2\to 1)})-\min(E_0^{(1\to 2)},E_1^{(1\to 2)})
-\min(E_0^{(2\to 1)},E_1^{(2\to 1)})\\
&=\Delta\hat{E}_\ell(h^{(1\to 2)},h^{(2\to 1)})=\min(h^{(1\to
2)},h^{(2\to 1)}).
\end{split}
\end{equation}
With the help of these equations, the distributions
$P_n^{(k)}(\Delta E)$ and $P_\ell(\Delta E)$ are easily obtained
once the distribution for the fields $P(h)$ is known.

Again similarly to the typical cavity method
\cite{MezardParisi01,MezardParisi03}, to derive the equation
satisfied by $P(h)$, we introduce $R^{(\gamma)}_{N+1}(h,E)$, the
probability to get an energy $E$ and cavity field $h$ when taking
at random a graph in $\G^{(\gamma)}_{N+1}$ and choosing one of its
node as a root. By definition, the marginalization over $h$ gives
$\mathbb{P}_{N+1}^{(\gamma)}(E/(N+1))$, the probability to get a
graph in $\G^{(\gamma)}_{N+1}$ with energy $E$,
\begin{equation}
\int dh R^{(\gamma)}_{N+1}(h,E)\equiv e^{(N+1)\L(E/(N+1),\gamma)}.
\end{equation}
Generalizing slightly Eq.~(\ref{eq:argument}), we write
\begin{equation}
\begin{split}
R^{(\gamma)}_{N+1}(h,E)&=\sum_{k=0}^\infty \pi_\gamma(k) \int
d\Delta E Q^{(k)}(h,\Delta E)e^{N\L[(E-\Delta
E)/N,\gamma-\chi(\gamma,k)/N]},\\
&\asymp e^{N\L(\e)}\sum_{k=0}^\infty \pi_\gamma(k) \int d\Delta
E Q^{(k)}(h,\Delta E)e^{-y\Delta E-z\chi(\gamma,k)},
\end{split}
\end{equation}
where $Q^{(k)}(h,\Delta E)$ denotes the joint distribution of the cavity field
$h$ and the energy shift $\Delta E$.
As in the typical cavity method, we verify that $h$ is
independent of $E$, more precisely,
\begin{equation}\label{eq:cmld}
\begin{split}
& R^{(\gamma)}_{N+1}(h,E)=e^{N\L
(\e)}e^{-\F(y,\gamma)}P(h),\\
& P(h_0)=\frac{1}{Z}\sum_{k=0}^\infty \pi_\gamma(k)  \int
\prod_{i=1}^kdh_iP(h_i)\delta(h_0-\hat{h}^{(k)}(\{h_i\}))e^{-y\Delta
\hat{E}_n^{(k)}(\{h_i\})-z\chi(\gamma,k)},\\
& Z = \sum_{k=0}^\infty \pi_\gamma(k) \int d\Delta E
P_n^{(k)}(\Delta E) e^{-y\Delta E}e^{-z\chi(\gamma,k)},\\
& \F(y,\gamma)=ye-\L(\e,\gamma)=-\ln Z,
\end{split}
\end{equation}
where $P(h)$ also depends on $\gamma$ and $y$. In the particular
case where $\chi(\gamma,k)$ does not depend on $k$, the relation
for $P(h)$ formally corresponds to what is known in the literature
as a 1RSB factorized equation with 1RSB parameter $y$ \cite{WongSherrington88,
MezardParisi03}; this is
the case with $\tilde{\G}^{(\gamma)}_N$ but not with
$\bar{\G}^{(\gamma)}_N$ and $\hat{\G}^{(\gamma)}_N$.

Specializing now to the vertex-cover problem, the equations are
simplified by the fact that $h\in\{0,1\}$, so that the
distribution $P(h)$ can be parameterized by a single real $\eta$,
with $P(h)=\eta\delta(h-1)+(1-\eta)\delta(h)$. As seen from
Eqs.~(\ref{eq:argument}) and (\ref{eq:z}), the distributions
$P_n^{(k)}(\Delta E)$ and $P_\ell(\Delta E)$ are needed only
through their Laplace transforms, which are given by
\begin{equation}
\begin{split}
&\int d\Delta E P_n^{(k)}(\Delta E)e^{-y\Delta E}=\int
\prod_{i=1}^k dh_i
P(h_i)e^{-y\Delta\hat{E}_n^{(k)}(\{h_i\})}=e^{-y}+(1-e^{-y})(1-\eta)^k,\\
&\int d\Delta E P_\ell(\Delta E)e^{-y\Delta E}=\int \prod_{i=1}^2
dh_i P(h_i)e^{-y\Delta\hat{E}_\ell(h_1,h_2)}=1+(e^{-y}-1)\eta^2.
\end{split}
\end{equation}
As a first check, one verifies that for $y=0$, the typical RS
ground-state energy density $\bar{\e}^{(\gamma)}$ is reobtained
\cite{WeigtHartmann00}, with same value for the three ensembles
$\tilde{\G}^{(\gamma)}_N$, $\bar{\G}^{(\gamma)}_N$ and
$\hat{\G}^{(\gamma)}_N$,
\begin{equation}\label{eq:eRS}
\begin{split}
&\bar{\e}^{(\gamma)}=1-\eta-\gamma\eta^2/2,\\
&\eta=e^{-\gamma\eta}.
\end{split}
\end{equation}

The equations for $y\neq 0$ can also be written explicitly. For
the ensemble $\tilde{\G}^{(\gamma)}_N$, they read
\begin{equation}
\begin{split}
&\eta=\frac{1}{1+e^{-y}(e^{\gamma \eta}-1)},\quad z=\frac{1}{2}(e^{-y}-1)\eta^2,\\
&\F(y,\gamma)=-\ln[e^{-y}+(1-e^{-y})e^{-\gamma
\eta}]+\frac{\gamma}{2}(e^{-y}-1)\eta^2.
\end{split}
\end{equation}
For the ensemble $\bar{\G}^{(\gamma)}_N$,
\begin{equation}\label{eq:same}
\begin{split}
&\eta=\frac{1}{1+e^{-y}(e^{\gamma \eta e^{-z}}-1)},\quad z=\ln[1+(e^{-y}-1)\eta^2],\\
&\F(y,\gamma)=-\ln [e^{-y}+(1-e^{-y})e^{-\gamma \eta
e^{-z}}]+\gamma (1-e^{-z})-\gamma z/2.
\end{split}
\end{equation}
And for the ensemble $\hat{\G}^{(\gamma)}_N$,
\begin{equation}
\begin{split}
&\eta=\frac{1}{1+e^{-y}(e^{\gamma \eta e^{-z}}-1)},\quad z=\frac{1}{2}\ln[1+(e^{-y}-1)\eta^2],\\
&\F(y,\gamma)=-\ln [e^{-y}+(1-e^{-y})e^{-\gamma \eta
e^{-z}}]+\gamma (1-e^{-z}).
\end{split}
\end{equation}
The formulae (\ref{eq:same}) coincide with the result of
the replica computation presented in \cite{MontanariZecchina02}.
\begin{figure}
\centering \epsfig{file=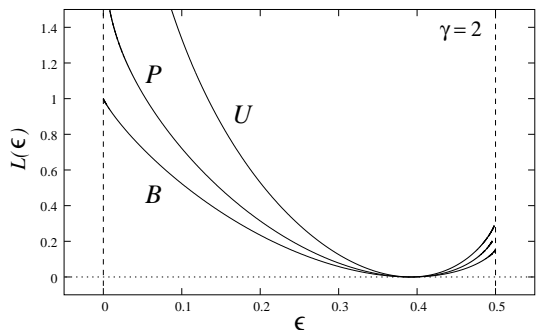,width=7cm}
\caption{Rate functions for the optimal energy $\e$ in the
vertex cover problem with the binomial model $\tilde{\G}^{(\gamma)}_N$ (label B),
the Poissonian model $\hat{\G}^{(\gamma)}_N$ (label P) and the uniform model
$\bar{\G}^{(\gamma)}_N$ (label U), all three for $\gamma=2$. The
common minimum at $\bar{\e}\simeq 0.39$ corresponds to the
prediction of the typical RS cavity method ($y=0$) \cite{WeigtHartmann00}. The larger
curvature of the rate function for the uniform model with respect
to the two other models can be interpreted by the fact that fixing
the ratio of edges imposes more constraints on the graph than
fixing the degree distribution or edge probability.\label{fig:vcg1}}
\end{figure}
The three corresponding rate functions are plotted in
Fig.~\ref{fig:vcg1} for $\gamma=2$.

A remarkable aspect of the vertex-cover problem is the presence,
in the typical phase diagram, of a continuous phase transition at
$\gamma_c=e\simeq 2.71$, from an RS phase at $\gamma<\gamma_c$ to
a presumably full-RSB phase at $\gamma>\gamma_c$
\cite{WeigtHartmann03b}. Due to its continuous character, the
phase transition can be located by analyzing the stability
analysis of the RS Ansatz. Extending the stability analysis from
typical to atypical graphs thus provides, in the $(\gamma,y)$
plane, a phase diagram with an "AT line" \cite{MezardParisi87b} separating a RS phase
from a full-RSB one. The three ensembles are not equivalent with
respect to properties associated with atypical graphs, and we
concentrate here on the binomial ensemble
$\tilde{\G}^{(\gamma)}_N$.

\begin{figure}
\begin{minipage}[t]{.46\linewidth}
\centering
\epsfig{file=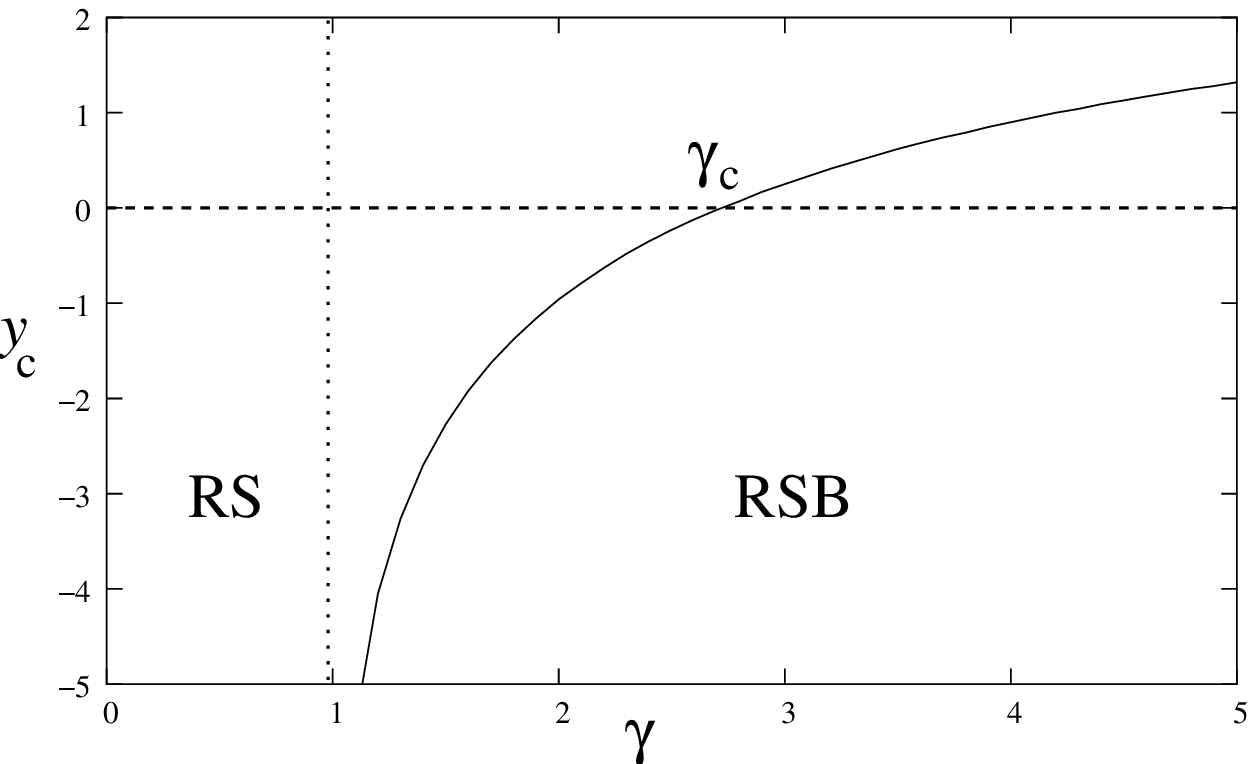,width=.8\linewidth}
\caption{Phase diagram of the vertex-cover problem in the
$(\gamma,y)$ plane for the Erd\H{o}s-R\'enyi ensemble $\tilde{\G}_N^{(\gamma)}$. 
The full line corresponds to the line
$y_c(\gamma)$ where the RS solution becomes instable ; there is no
instability below the percolation threshold, 
$\gamma<1$ [$y_c(\gamma)\to-\infty$ for $\gamma\to 1^+$]. The line $y=0$
reproduces the phase diagram of the typical case with
$\gamma_c=e\simeq 2.71$ being defined as the intersection of
$y_c(\gamma)$ with $y=0$. \label{fig:vcdiag}}
\end{minipage} \hfill
\begin{minipage}[t]{.46\linewidth}
\centering\epsfig{file=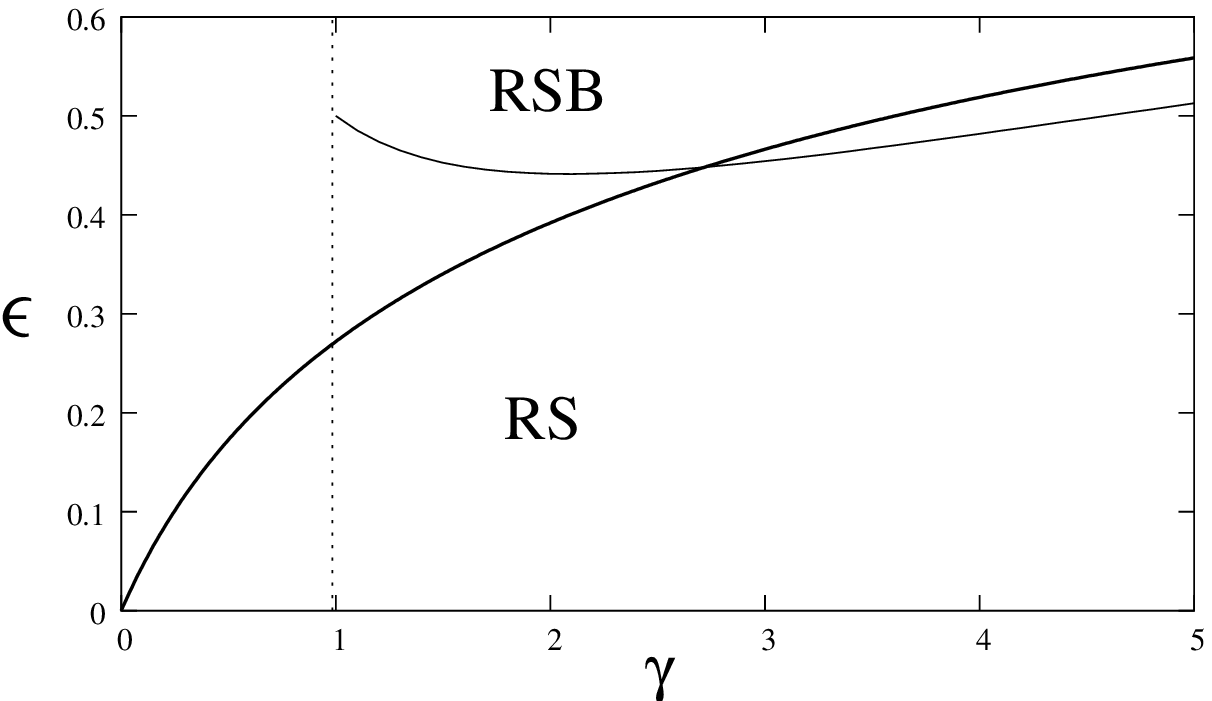,width=.8\linewidth}
\caption{Phase diagram of the vertex-cover problem in the $(\gamma,\e)$ plane for 
the Erd\H{o}s-R\'enyi ensemble $\tilde{\G}_N^{(\gamma)}$.
The full line represents the typical RS energy density, as given by
Eq.~(\ref{eq:eRS}), and the thin line, starting only from $\gamma=1$, the energy density above
which the RS approach fails, as given by Eq.~(\ref{eq:stabRS}).
The two curves cross at $\gamma_c=e$, corresponding to the onset of RSB on typical graphs.\label{fig:vcstab}}
\end{minipage}
\end{figure}

RSB effects are much likely to appear first for negative values of
$y$, corresponding to the most frustrated graphs. The failure of
the RS approach can in fact be inferred from an asymptotic
analysis of the $y\to-\infty$ limit : it yields $\eta\sim
e^{y/2}/\sqrt{\gamma}$, $\e(y=-\infty)=1/2$ and
$L(y=-\infty)=(1-\ln\gamma)/2$. Clearly, this is inconsistent as
soon as $\gamma>e$ since then it predicts then $L(\e=1/2)<0$. The
value thus obtained coincides with the value of $\gamma_c$
for the failure of the RS approach to typical
graphs \cite{WeigtHartmann00, BauerGolinelli01} (the reason for this correspondence
is elucidated below). The negativeness of the rate
function is however a sufficient but not necessary sign of RSB. A
more refined way to detect it consists in studying the stability of
the RS large deviations Ansatz. For the binomial model, it happens
to be strictly equivalent to the stability analysis of a
factorized 1RSB Ansatz \cite{MontanariRicci03}, and reads
\begin{equation}\label{eq:stabRS}
(\gamma \eta)^2e^{-y}<1.
\end{equation}
It starts to be violated at $y=-\infty$ for $\gamma>1$, while
the typical graphs described with $y=0$ are not concerned before
$\gamma=e$. Indeed for $\gamma<1$, the RS Ansatz is stable for
all $y$: $(\gamma \eta)^2e^{-y}$ is a decreasing function of $y$
and for $y\to-\infty$ the asymptotic analysis yields $(\gamma
\eta)^2e^{-y}\sim \gamma$. At $\gamma=1$, only the $y=-\infty$
point, corresponding to the maximum achievable energy $\e=1/2$, is
marginally unstable. Finally, for $\gamma>1$, there is a critical
$y_c$ such that RS is stable for $y>y_c$ but unstable for $y<y_c$ ;
$y_c$ increases when $\gamma$ increases and reaches $y_c=0$ for
$\gamma_c=e$, the point where the typical problem undergoes the
RSB transition. For $\gamma>\gamma_c$ while typical graphs are
FRSB, some less frustrated graphs are still RS. The resulting
phase diagram is shown in Fig.~\ref{fig:vcdiag} in the plane
$(\gamma,y)$ and in Fig.~\ref{fig:vcstab} in the plane
$(\gamma,\e)$. The occurrence of RSB at $\gamma=1$ is particularly
interesting because this point corresponds to the percolation
threshold of a giant connected component \cite{Bollobas01}, which appeared totally
irrelevant when restricting to
typical graphs \cite{WeigtHartmann00}. In contrast, when atypical
graphs are included into the picture, the emergence of a giant
component seems to be responsible for the onset of RSB, as shown
in Fig.~\ref{fig:vcdiag} (a similar analysis of the uniform model
however reveals that in this case RSB appears only above an
average connectivity $\gamma=2$).

The opposite $y\to+\infty$ limit is also interesting since it is
always correctly described by the RS Ansatz, with
$\e(y=\infty,\gamma)=0$ and
$\mathcal{L}(y=\infty,\gamma)=-\gamma/2$. It can be interpreted as
corresponding to graphs with no edge at all, which occurs with
probability
\begin{equation}
\P_N^{(\gamma)}({\rm non\ frustrated})\asymp\P_N^{(\gamma)}({\rm
no\ edge})\asymp \left(1-\frac{\gamma}{N}\right)^{N^2/2}\asymp
e^{-N\gamma/2}.
\end{equation}
Similar relations between the $y=\infty$ limit and the probability
for non-frustrated samples have been reported in a variety of other
models \cite{Rivoire04}, providing consistent checks of the
method.

\subsection{Random graphs with given degree distributions}\label{sec:graphreg}

The LDCM applies as well to graph
ensembles with non-poissonian degrees, and an example is provided
here with graph ensembles specified by their degree distribution
$p(k)$, that is with each node having an independent probability $p(k)$ of
being of degree $k$. The reasoning for arbitrary $p(k)$ basically
follows the procedure used for the Poissonian model
$\hat{\G}_N^{(\gamma)}$ which was a particularly case where
$p(k)=\gamma^ke^{-\gamma}/k!$.

We thus first consider how the ensemble is modified when a new
node is connected to $k$ nodes of a graph made of $N$ nodes having
degree distribution $p(K)$. The degree distribution becomes $p'(K)$, with
\begin{equation}\label{eq:k}
p'(K)=\left(1-\frac{k}{N}\right)p(K)+\frac{k}{N}p(K-1),
\end{equation}
since a given node has a probability $k/N$ to get its degree
increased by one unit (the probability that it is increased by
more than one unit is $O(1/N^2)$ and is therefore neglected).
Writing $p(K)=\sum_r p_r\delta_{K,r}$, we explicitly have
$p_r'=p_r+(k/N)\delta p_r$ with $\delta p_r=p_{r+1}-p_r$. The set $\{p_r\}$ can serve as a characterization of the graph
ensemble, and following the same scheme as for Poissonian graphs, we
obtain
\begin{equation}
e^{-\F(y,\{p_r\})}=\sum_{k=0}^\infty p_k\int d \Delta E
P_n^{(k)}(\Delta E)e^{-y\Delta E}e^{-kz},
\end{equation}
with as before $y=\partial_\e \L(\e,\{p_r\})$ and now
\begin{equation}
z\equiv\sum_r \delta p_r\frac{\partial\L(\e,\{p_r\})}{\partial
p_r}.
\end{equation}
To get $z$, we notice that Eq.~(\ref{eq:k}) with $k=2$ describes
the effect of an edge addition so that
\begin{equation}
e^{2z}=\int d\Delta E P_\ell(\Delta E)e^{-y\Delta E}.
\end{equation}
The potential whose Legendre transform yields the rate function
can therefore be written
\begin{equation}\label{eq:pot}
\F(y,\{p_r\})=-\ln\left[\sum_{k=0}^\infty p_k\int d\Delta E
P_n^{(k)}(\Delta E)e^{-y\Delta E}\left(\int d\Delta E'
P_\ell(\Delta E')e^{-y\Delta E'}\right)^{-k/2}\right].
\end{equation}
Similarly, the cavity equation for the fields reads
\begin{equation}\label{eq:caveq}
P(h_0)=\frac{1}{Z}\sum_{k=0}^\infty \frac{(k+1) p_{k+1}}{\langle
k\rangle}\int
\prod_{j=1}^kdh_jP(h_j)\delta(h_0-\hat{h}^{(k)}(\{h_j\}))e^{-y\Delta
\hat{E}^{(k)}_n(\{h_j\})}e^{-kz}
\end{equation}
where $Z$ is the appropriate normalization and the presence of
$(k+1)p_{k+1}/\langle k\rangle$ instead of $p_k$ reflects the fact
that an oriented edge is chosen at random, and not a node as in Eq.~(\ref{eq:pot}) (only for Poissonian graphs do these two
probabilities happen to be the same).

A case of particular interest is when $p(k)=\delta_{k,r}$,
corresponding to random $r$-regular graphs. In such a case, the
factor $e^{-kz}$ in Eq.~(\ref{eq:caveq}) is a constant which can be
absorbed into the normalization $Z'=Ze^{(r-1)z}$,
\begin{equation}
P(h_0)=\frac{1}{Z'}\int
\prod_{j=1}^{r-1}dh_jP(h_j)\delta(h_0-\hat{h}^{(r-1)}(\{h_j\}))e^{-y\Delta
\hat{E}^{(r-1)}_n(\{h_j\})}.
\end{equation}
This is formally identical to what is known as a 1RSB
factorized cavity equation \cite{WongSherrington88,MezardParisi03}. The correspondence extends to the
formula for the potential, Eq.~(\ref{eq:pot}), which becomes for
regular graphs,
\begin{equation}
\F(y)=-\ln\left[\int d\Delta E P_n^{(r)}(\Delta
E)e^{-y\Delta E}\right]+\frac{r}{2}\ln\left[\int d\Delta E
P_\ell(\Delta E)e^{-y\Delta E}\right].
\end{equation}
As a consequence, the 1RSB complexity of models defined on random
regular graphs coincides with minus a rate function, as already
noticed in \cite{Rivoire04}. Obviously, the correspondence holds
only within the RS approximation that has been assumed throughout
since by nature of a 1RSB glassy phase, the complexity is positive
while a rate function is necessarily non-negative ; it will be
shown below how the formalism needs to be extended to include the possibility
of RSB. When the factorization does not hold, the correspondence
between rate functions over the external disorder and negative
complexities is only approximate ; we have seen however that for
vertex-cover on $\tilde{\G}^{(\gamma)}_N$ the rate function starts
getting negative values precisely at the point $\gamma=e$ where
typical graphs undergo a RSB transition, in agreement with the
observation that the two quantities are approximatively related.

\section{Multi-step large deviations}

The LDCM can naturally be extended
beyond the simple case of zero-temperature systems in an RS phase
with disorder only specified by a random graph ensemble. We
consider here successively finite-temperature systems,
models with RSB phases, and external disorders including random
couplings. In the three cases, a second temperature is needed to
describe the large deviations with respect to the additional
source of randomness. In each case also, the equations display a
common 2RSB-like structure \cite{MezardParisi87b}, which would be
promoted to the $n$RSB type with $n>2$ if $n$ different sources of 
disorder were present.

\subsection{Finite temperature}\label{sec:finiteT}

The simplest extension requiring multi-step large deviations
consists in generalizing the description of a
model on a given graph from zero to finite temperature. Two inverse temperatures are now required : $\beta$,
for the thermodynamics on a given graph, and $y$ for the large
deviations in the graph ensemble. More precisely, large deviations
now concern the density of free energy $f(\beta)$, with the limit
$\beta\to\infty$ giving back to the large deviations for the
ground-state density energy $\e=\lim_{\beta\to\infty}f(\beta)$, as
discussed so far. For any fixed value of $\beta$, the rate
function $L(f,\beta)\equiv -\L(f,\beta)$ is calculated as before through the Legendre
transform of a potential $\F(y,\beta)$ satisfying
\begin{equation}\label{eq:finiteT}
\begin{split}
&e^{-N\F(y,\beta)}=\int df\ e^{N[\L(f,\beta)-yf]}=\frac{1}{\#\G}\sum_{G\in\mathcal{G}}Z_G(\beta)^{y/\beta},\\
&Z_G(\beta)\equiv e^{-\beta
Nf_G(\beta)}\equiv\sum_{C\in\mathcal{C}_G}e^{-\beta E(C)},
\end{split}
\end{equation}
where we introduced $Z_G(\beta)$ the partition function on the
graph $G$ at temperature $\beta$, $\mathcal{C}_G$ the set of
configurations on the graph $G$ and $\#\G$ the cardinality of the
ensemble of graphs $\G$. Note that the particular choice $y=\beta$
corresponds to the uniform measure over all configurations
$\{C\in\mathcal{C}_G\}_{G\in\mathcal{G}}$ :
\begin{equation}
\sum_{G\in\mathcal{G}}\sum_{C\in\mathcal{C}_G}e^{-\beta
E(C)}=\sum_{G\in\mathcal{G}}e^{-\beta N f_G(\beta)}=(\#\G) e^{-\beta
N\F(\beta,\beta)}.
\end{equation}

From the technical point of view, we just have to replace in all formulae the
functions $\hat{h}^{(k)}(\{h_i\})$, $\Delta\hat{E}_n^{(k)}(\{h_i\})$ and
$\Delta\hat{E}_\ell(h_1,h_2)$ by their
finite-temperature extensions $\hat{h}^{(k)}(\{h_i\};\beta)$,
$\Delta\hat{F}_n^{(k)}(\{h_i\};\beta)$ and
$\Delta\hat{F}_\ell(h_1,h_2;\beta)$. Taking the vertex-cover problem
as an example, these quantities are derived by writing recursive
equations for the conditional partition functions $Z_0^{(i\to
0)}(\beta)$ and $Z_1^{(i\to 0)}(\beta)$ instead of the
conditional ground-state energies $E_0^{(i\to 0)}$ and $E_1^{(i\to
0)}$. More precisely, Eqs.~(\ref{eq:basicvc}) are replaced with
\begin{equation}
\begin{split}
Z_0^{(i\to 0)}(\beta) = &e^{-\beta}\prod_{j=1}^k\left(Z_0^{(j\to
i)}(\beta)+Z_1^{(j\to i)}(\beta)\right),\\
Z_1^{(i\to 0)}(\beta) = &\prod_{j=1}^k Z_0^{(j\to i)}(\beta).
\end{split}
\end{equation}
To get $\lim_{\beta\to\infty}h^{(i\to 0)}(\beta)=h^{(i\to 0)}$
with $h^{(i\to 0)}$ defined in Eq.~(\ref{eq:fieldsdef}), the
cavity fields at finite temperature $h^{(i\to 0)}(\beta)$ are
defined as
\begin{equation}
h^{(i\to 0)}(\beta)\equiv -\frac{1}{\beta}\ln\left(\frac{Z_0^{(i\to
0)}(\beta)}{Z_0^{(i\to 0)}(\beta)+Z_1^{(i\to 0)}(\beta)}\right).
\end{equation}
With these definitions, the different functions required to
compute the rate function $L(f,\beta)$ are
\begin{equation}
\begin{split}
&\hat{h}^{(k)}(\{h_j\},\beta)=\frac{1}{\beta}\ln\left(1+e^{\beta(1-\sum_{j=1}^kh_j)}\right),\\
&\Delta\hat{F}_n^{(k)}(\{h_j\};\beta)=-\frac{1}{\beta}\ln\left(e^{-\beta}+e^{-\beta\sum_{j=1}^kh_j}\right),\\
&\Delta\hat{F}_\ell(h_1,h_2;\beta)=-\frac{1}{\beta}\ln\left(e^{-\beta
h_1}+e^{-\beta h_2}-e^{-\beta (h_1+h_2)}\right),
\end{split}
\end{equation}
which all reduce as it should to $\hat{h}^{(k)}(\{h_j\})$,
$\Delta\hat{E}_n^{(k)}(\{h_j\})$ and $\Delta\hat{E}_\ell(h_1,h_2)$
given in Eq.~(\ref{eq:fieldsdef}), (\ref{eq:Enode}) and
(\ref{eq:Eedge}) when $\beta\to\infty$. The only practical
difference with the $\beta=\infty$ case is that the distribution
$P(h)$ has no more reason to be peaked on integers and therefore
cannot be parameterized by a single real number.

\subsection{Replica symmetry breaking}\label{sec:rsb}

The phase space of a glassy 1RSB instance is structured into clusters
whose energy is controlled by a parameter $\mu$ in exactly the
same way the finite inverse temperature $\beta$ controls the
equilibrium configurations according to their energy. Therefore,
the extension of the LDCM from
paramagnetic (RS) systems to glassy (1RSB) systems, is formally
similar to the extension from zero temperature to finite
temperature. The counterpart of Eq.~(\ref{eq:finiteT}) reads
\begin{equation}\label{eq:rsbpot}
\begin{split}
&e^{-N\F(y,\mu)}=\frac{1}{\#\G}\sum_{G\in\mathcal{G}}e^{-yN\phi_G(\mu)}=\int 
d\phi\ e^{N[\L(\phi,\mu)-y\phi]},\\
&e^{-N\mu\phi_G(\mu)}=\sum_{\alpha\in G}e^{-\mu N
\e_\alpha}=\int d\e\ e^{N[\Sigma_G(\e)-\mu \e]}
\end{split}
\end{equation}
where $\phi_G(\mu)$ is the 1RSB potential on graph $G$ and
$L(\phi,\mu)\equiv -\L(\phi,\mu)$ is the rate function describing the large deviations of $\phi(\mu)$ over the ensemble of random graphs ; in these
formulae, we reserve greek letters for quantities related to the
internal structure and use $\alpha$ to index the clusters. The
saddle points in Eq.~(\ref{eq:rsbpot}) lead to the following
Legendre transform relations :
\begin{eqnarray}\label{eq:pot2}
\F(y,\mu)&=y\phi-\L(\phi,\mu),& y=\partial_{\phi}\L(\phi,\mu),\\
\mu\phi(\mu)&=\mu \e -\Sigma(\e),&\mu =\partial_\e\Sigma(\e).\\
\end{eqnarray}
These quantities are computed by applying the standard 1RSB cavity
method \cite{MezardParisi01} to a given set of atypical graphs characterized by their
ground-state energy density $\e_0$. If $\rho_N(\epsilon|\e_0)$ denotes the
distribution of their clusters, the corresponding complexity is
defined as $\rho_N(\epsilon|\e_0)\asymp\exp [N
\Sigma(\epsilon|\e_0)]$ ; for an energy $\epsilon$ close to the
ground-state reference energy $\e_0$, it becomes
\begin{equation}
\rho_N(\epsilon|\e_0)\asymp e^{\mu N(\epsilon-\e_0)}
\end{equation}
where $\mu\equiv\partial_\epsilon\Sigma(\epsilon=\e_0|\e_0)$
defines the 1RSB internal inverse temperature. The shift in energy $\Delta E$ induced
by a node addition, which is needed in the recursion at the level of the
graph average, is given by the shift in the reference energy,
that is,
\begin{equation}
\rho_{N+1}(\epsilon|\e_0)= \rho_N(\epsilon|\e_0)e^{-\mu\Delta E}.
\end{equation}
The expression for the reweighting term $\Xi\equiv e^{-\mu\Delta E}$ is read from the 1RSB cavity recursion which involves $\Pi(h)$, the distribution of
cavity fields over the clusters, and $P[\Pi]$, the distribution of the
$\Pi$'s over the graphs ; for a given (class of) graph, the
connection of a new node to $k$ other ones indeed yields
\begin{equation}\label{eq:rsb}
\begin{split}
\Pi_0=\hat{\Pi}^{(k)}[\{\Pi_i\}],\quad {\rm with}\quad &
\hat{\Pi}^{(k)}[\{\Pi_i\}](h_0)=\frac{1}{\Xi}\int\prod_{i=1}^k\Pi_i(h_i)\delta(h_0-\hat{h}^{(k)}(\{h_i\}))e^{-\mu\Delta\hat{E}_n^{(k)}(\{h_i\})},\\
&\Xi=e^{-\mu\Delta
E}=\hat{\Xi}^{(k)}[\{\Pi_i\}]=\int\prod_{i=1}^k\Pi_i(h_i)e^{-\mu\Delta\hat{E}_n^{(k)}(\{h_i\})}.
\end{split}
\end{equation}

Therefore, at the level of the graph average, we have for Poissonian graphs
\begin{equation}\label{eq:rsb2}
P[\Pi_0]=\frac{1}{Z}\sum_{k=0}^\infty
\pi_\gamma(k)\int\prod_{i=1}^k\D \Pi_i
P[\Pi_i]\ \delta[\Pi_0-\hat{\Pi}^{(k)}[\{\Pi_i\}]\ \hat{\Xi}^{(k)}(\{\Pi_i\})]^{y/\mu}\ e^{-z\chi(k,\gamma)}, 
\end{equation}
and for graphs with fixed degree distribution
\begin{equation}
P[\Pi_0]=\frac{1}{Z'}\sum_{k=0}^\infty \frac{(k+1)
p_{k+1}}{\langle k\rangle}\int
\prod_{i=1}^k\D\Pi_iP[\Pi_i]\ \delta[\Pi_0-\hat{\Pi}^{(k)}(\{\Pi_i\})]\ \hat{\Xi}^{(k)}[\{\Pi_i\}]^{y/\mu}\ e^{-kz}.
\end{equation}
The 1RSB large deviations equations have thus the structure of a typical
factorized 2RSB theory, as the RS large deviations equations
resembled a typical factorized 1RSB theory. In particular, for
$y=0$, the non-factorized 1RSB formalism is exactly recovered.
\label{sec:coloring}

Replica symmetry breaking (RSB) is relevant to many optimization
problems, and the vertex-cover problem already provided us with
such an example. For this problem, studying the local stability of
the RS Ansatz was enough to locate the continuous transition to a
RSB phase. However, other problems may display a different kind of
glass transition, known as a discontinuous 1RSB transition,
which, due to its discontinuous character, can only be correctly
described by implementing a 1RSB formalism. Such a transition is
found for instance in the coloring problem \cite{MuletPagnani02},
which we take here as an illustrative example of the broader class
of constraint satisfaction problems.

In the statistical physics point of view, a problem is satisfiable (SAT) if
it has a zero ground-state energy density, $\e=0$~\cite{footnoteZero}. In
presence of a clustered glassy phase however, an alternative
characterization is provided by the sign of the complexity
$\Sigma_0$, giving, when positive, the number $\exp[N\Sigma_0]$ of
clusters with $\e=0$. This complexity $\Sigma_0$ is obtained in
the 1RSB formalism by taking the limit $\mu\to\infty$ ; for
the 3-coloring problem on Erd\H{o}s-R\'enyi graphs
$\tilde{\G}_N^{(\gamma)}$, $\Sigma_0$ is
found to be positive only in a restricted interval $[\gamma_d,\gamma_c]$,
with $\gamma_d=4.42$ and $\gamma_c=4.69$~\cite{MuletPagnani02}, as schematically 
represented in Fig.~\ref{fig:cartoon}. The threshold values $\gamma_d$ and
$\gamma_c$, which also appear in other constraint satisfaction problems
such as $K$-SAT~\cite{MezardZecchina02}, locate two phase
transitions, called respectively the clustering and SAT-UNSAT
transitions : with probability one when $N\to\infty$, a graph with
$\gamma<\gamma_d$ is colorable and RS, a graph with
$\gamma_d<\gamma<\gamma_c$ is again colorable but RSB, and a graph
with $\gamma>\gamma_c$ is uncolorable.

As an illustration, we consider here the 3-coloring problem on the Erd\H{o}s-R\'enyi 
ensemble $\tilde{\G}_N^{(\gamma)}$. Following~\cite{BraunsteinMulet03}, Eq.~(53), the shift $\Delta\phi_\ell$ in the
1RSB potential due to a link addition is given by
\begin{equation}
e^{-\mu\Delta \phi_\ell}=1+(e^{-\mu}-1)q\eta_1\eta_2,
\end{equation}
where the $\eta_j$'s, with distribution $\rho(\eta)$, represent the
1RSB cavity fields for this problem~\cite{BraunsteinMulet03}.
In the LDCM, we need the Laplace transform of the distribution $P_\ell(\Delta\phi)$,
which thus reads
\begin{equation}
\int d\Delta\phi P_\ell(\Delta\phi)e^{-y\Delta \phi}=\int
\prod_{i=1,2}d\eta_i\rho(\eta_i)\left(1+(e^{-\mu}-1)q\eta_1\eta_2\right)^{y/\mu}.
\end{equation}
If only SAT configurations are to be addressed, the
general 1RSB-LDCM equations can be simplified by taking
the limit $\mu\to\infty$. This limit enforces $\e\to 0$ and
$\mu\phi(\mu)\to -\Sigma_0$ and requires to rescale $y$ by taking
$y\to\infty$ with $x=y/\mu$ fixed, such that $\F(y,\mu)\to \F(x)$
with
\begin{equation}
\F(x)=x\Sigma_0-\L(\Sigma_0),\quad
x=\partial_{\Sigma_0}\L(\Sigma_0),
\end{equation}
where $\L(\Sigma_0)=\lim_{\mu\to\infty}\L(\phi=-\Sigma_0/\mu,\mu)$.
In this limit,
\begin{equation}
\int d\Delta\phi P_\ell(\Delta\phi)e^{-y\Delta \phi}\to\int
\prod_{i=1,2}d\eta_i\rho(\eta_i)\left(1-q\eta_1\eta_2\right)^x.
\end{equation}
Similarly for site addition, we have, again in the limit $\mu\to\infty$,
\begin{equation}
\int d\Delta\phi P_n^{(k)}(\Delta\phi)e^{-y\Delta \phi}\to\int
\prod_{i=1}^kd\eta_i\rho(\eta_i)\hat{\Xi}^{(k)}(\eta_1,\dots,\eta_k)^x,
\end{equation}
with
\begin{equation}
\Xi^{(k)}\equiv\lim_{\mu\to\infty}e^{-\mu\Delta
\phi^{(k)}_n}=\sum_{\ell=0}^{q-1}(-1)^\ell\binom{q}{\ell+1}\prod_{i=1}^k\left(1-(\ell+1)\eta_i\right),
\end{equation}
where $\Delta\phi^{(k)}_n$ refers to the shift in potential due to the
connection of a new nodes to $k$ old ones (see Eq.~(56) in~\cite{BraunsteinMulet03}).

\begin{figure}
\begin{minipage}[t]{.3\linewidth}
\centering
\epsfig{file=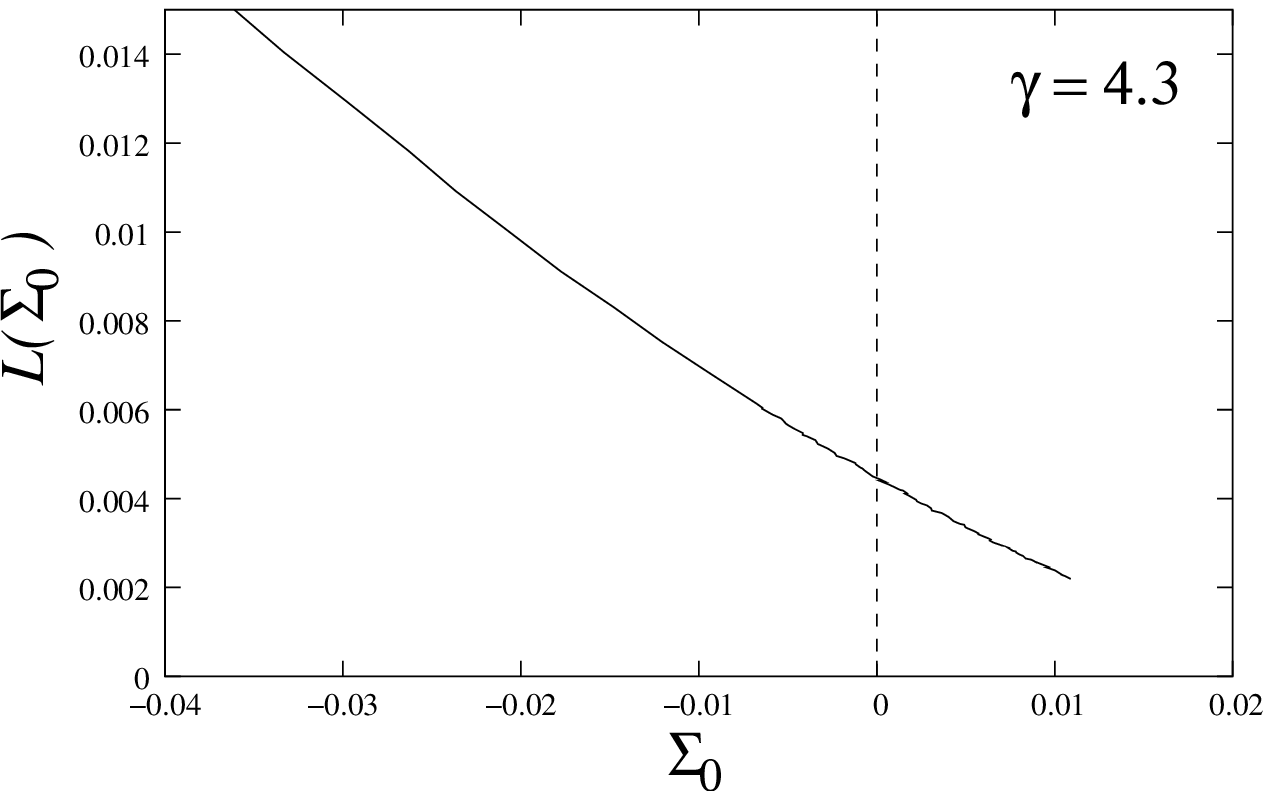,width=\linewidth}
\end{minipage} \hfill
\begin{minipage}[t]{.3\linewidth}
\centering
\epsfig{file=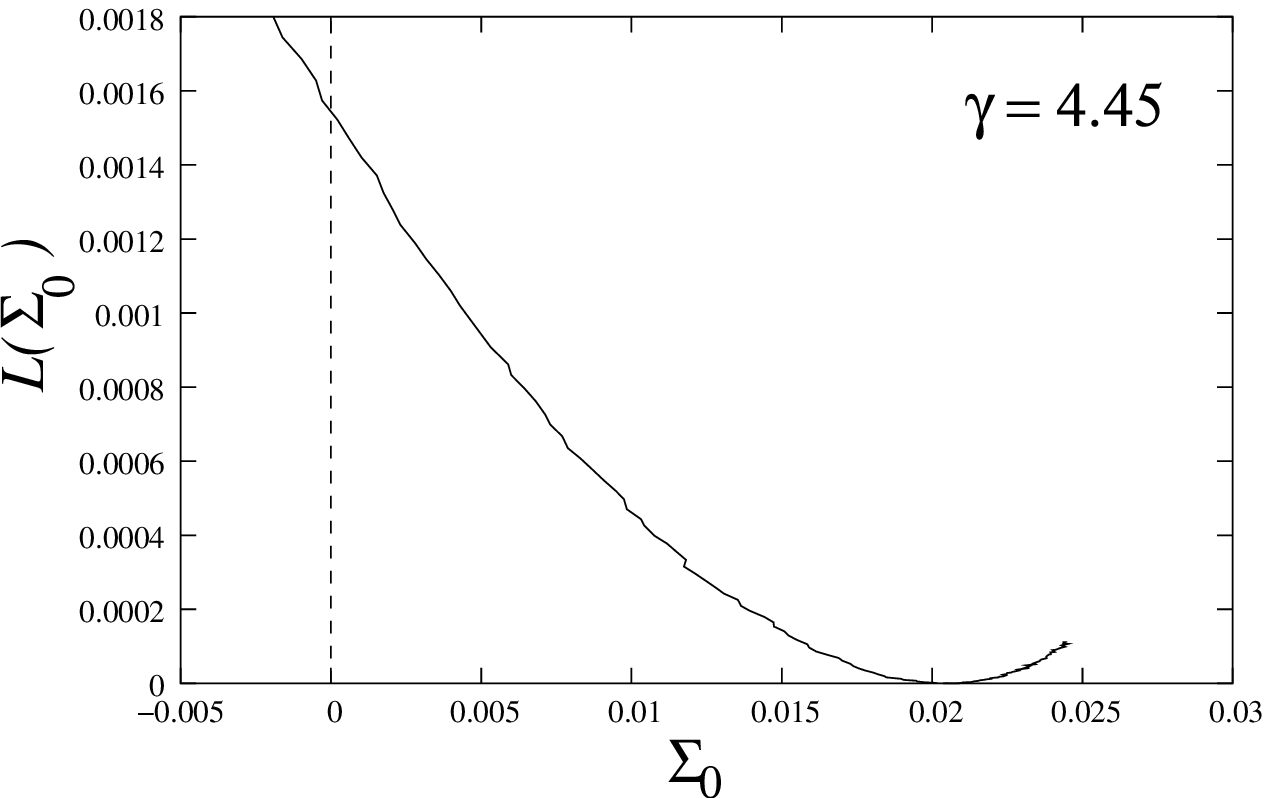,width=\linewidth}
\end{minipage} \hfill
\begin{minipage}[t]{.3\linewidth}
\centering
\epsfig{file=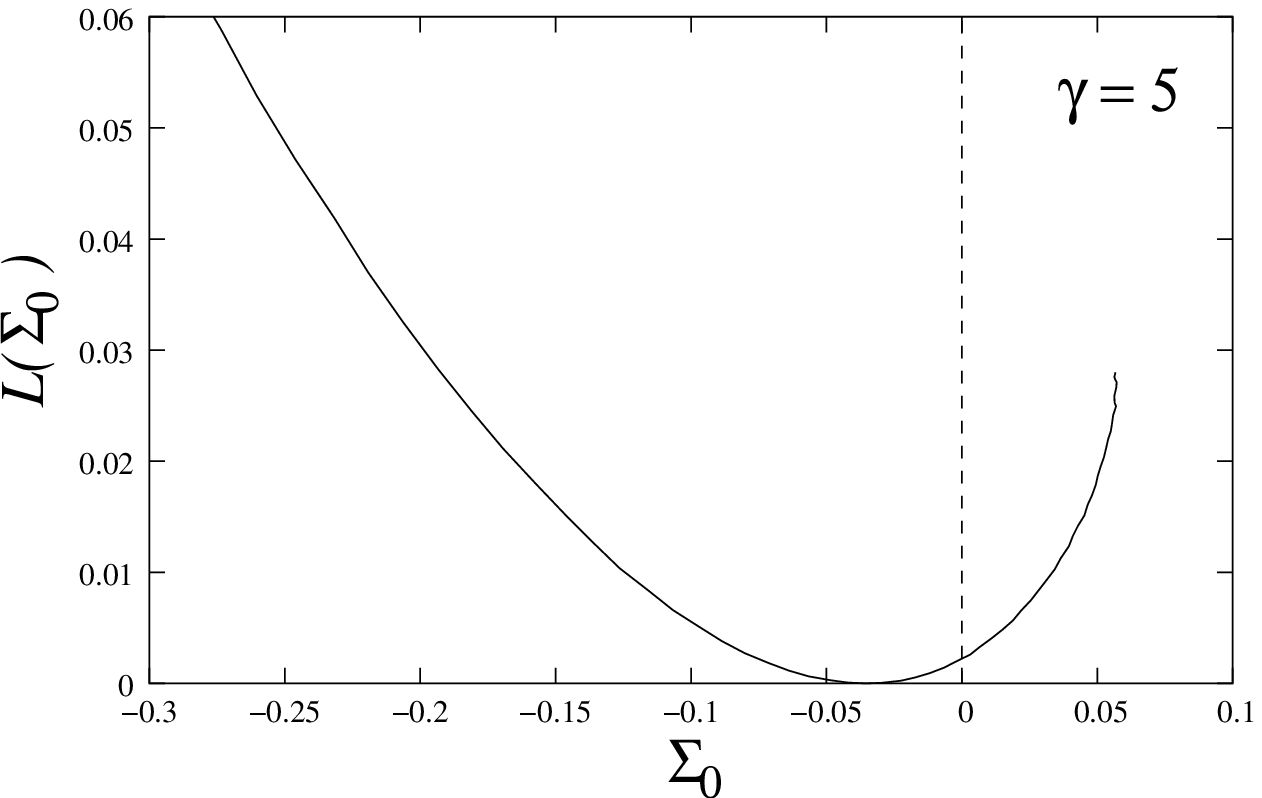,width=\linewidth}
\end{minipage}
\caption{Rate functions $L(\Sigma_0)$ of the complexity
$\Sigma_0$ in the 3-coloring problem for the ensemble
$\tilde{\G}_N^{(\gamma)}$ with $\gamma=4.3<\gamma_d$,
$\gamma=4.45\in[\gamma_d,\gamma_c]$ and
$\gamma=5>\gamma_c$, as obtained from the 1RSB-LDCM. A maximal value $x_d$ of the slope $x=-\partial_{\Sigma_0} L(\Sigma_0)$, associated with a maximal value of $\Sigma_0$, is found above which a non-trivial $L(\Sigma_0)$ ceases
to exist with, by definition of $\gamma_d$, $x_d(\gamma)<0$ when $\gamma<\gamma_d$, and $x_d(\gamma)>0$ when $\gamma>\gamma_d$. Note that for $\gamma=4.3$, the curve displayed certainly does not describe the actual rate function which is expected to
vanish at $\Sigma_0=0$ : since the non-trivial solution shown coexists with the trivial solution reduced to the point $(\Sigma_0=0$, $L=0)$, the correct rate function may be obtained by a Maxwell construction i.e., by drawing the supporting line originating from $(\Sigma_0=0$, $L=0$) and tangent to the curve.
\label{fig:ldfcol}}
\end{figure}

\begin{figure}
\begin{minipage}[t]{.46\linewidth}
\centering
\epsfig{file=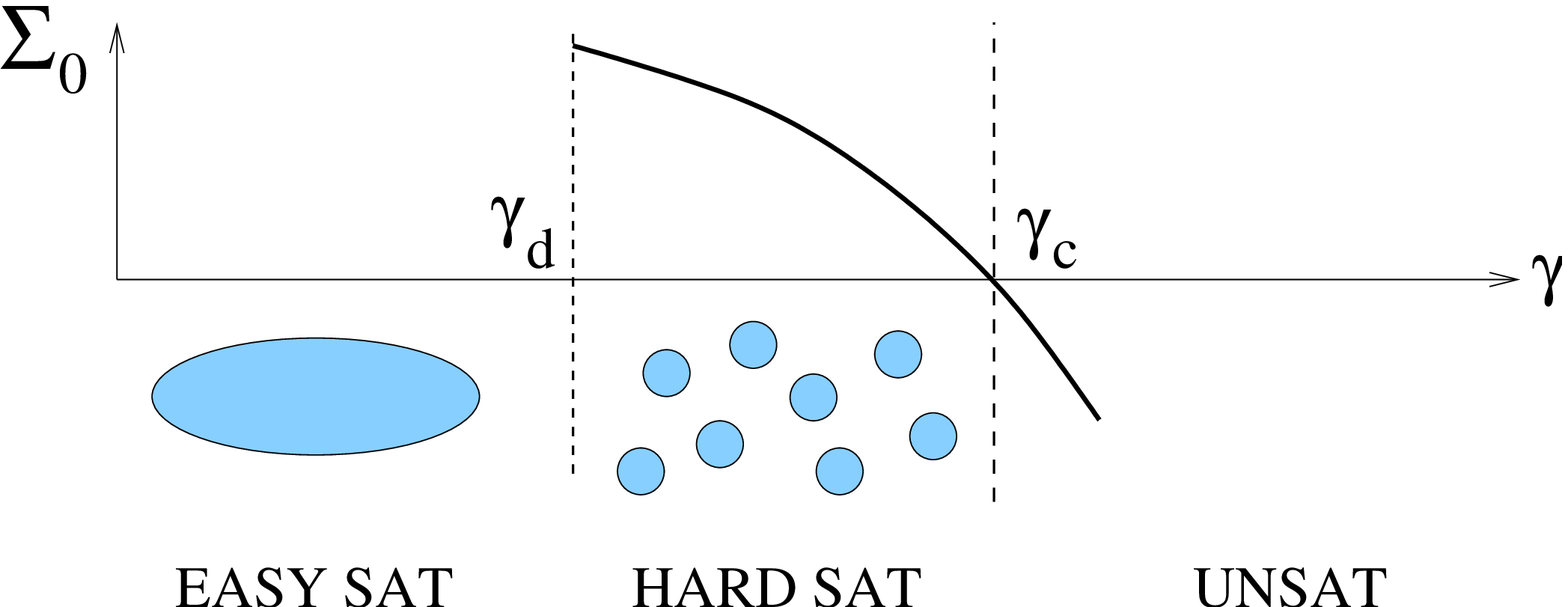,width=\linewidth}
\caption{Pictorial representation of the phase diagram for
typical instances of the 3-coloring problem on Erd\H{o}s-R\'enyi
graphs $\tilde{\G}_N^{(\gamma)}$. For $\gamma<\gamma_d$ the set of solutions forms a single
connected cluster. For $\gamma_d<\gamma<\gamma_c$, the set of
solutions is organized into  $\exp (N\Sigma_0)$ clusters, where
$\Sigma_0$ is the complexity curve represented in the upper part.
For $\gamma>\gamma_c$, $\Sigma_0<0$, indicating that there is
(typically) no solution, i.e., the system is UNSAT.\label{fig:cartoon}}
\end{minipage} \hfill
\begin{minipage}[t]{.46\linewidth}
\centering
\epsfig{file=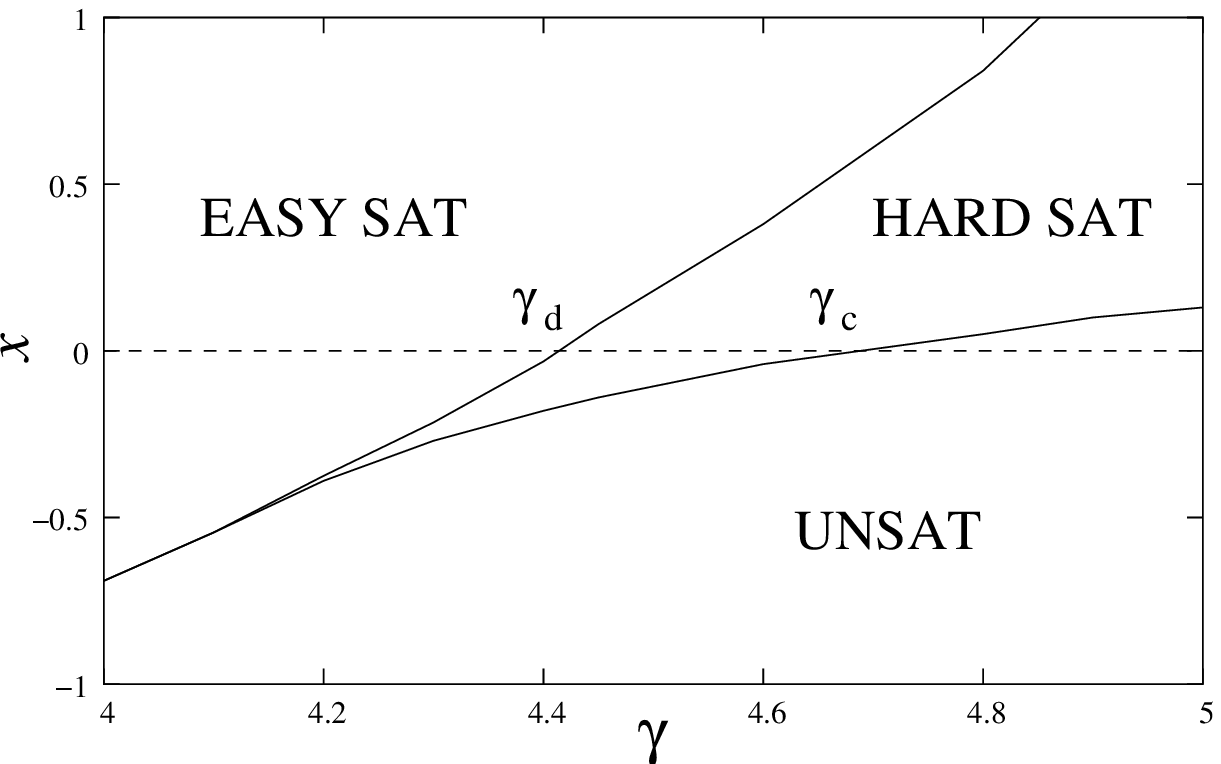,width=.8\linewidth}
\caption{Phase diagram of the 3-coloring problem on
Erd\H{o}s-R\'enyi graphs in the $(\gamma,x)$ plane. The top curve
gives $x_d(\gamma)$, the value of $x$ at which a 1RSB solution
appears and the bottom curve gives $x_c(\gamma)$, the value of $x$
at which the complexity vanishes. The intersection of these two
curves with the line $x=0$ gives back the typical phase diagram of Fig.~\ref{fig:cartoon} with the intersection points $\gamma_d$ and $\gamma_c$. (Note
that this diagram includes values from non-physical branches as 
discussed  in Fig.~\ref{fig:ldfcol} for $\gamma=4.3$.)\label{fig:phasediag}}
\end{minipage}
\end{figure}

The distribution $\rho(\eta)$ is determined, in the limit
$\mu\to\infty$, by the self-consistent equation
\begin{equation}
\rho(\eta_0)=\frac{1}{Z}\sum_{k=0}^\infty
\pi_\gamma(k)\int\prod_{i=1}^kd\eta_i\rho(\eta_i)\ \delta(\eta_0-\hat{\eta}^{(k)}(\{\eta_i\}))\ \hat{\Xi}^{(k)}(\{\eta_i\})^x\ e^{-z\chi(k,\gamma)},
\end{equation}
with
\begin{equation}
\hat{\eta}^{(k)}(\{\eta_i\}))\equiv\frac{1}{\hat{\Xi}^{(k)}(\{\eta_i\})}\sum_{\ell=0}^{q-1}(-1)^\ell\binom{q-1}{\ell}\prod_{i=1}^k\left(1-(\ell+1)\eta_i\right).
\end{equation}
This equation can be solved numerically using a population
dynamics algorithm~\cite{MezardParisi01} : as for the typical
case~\cite{BraunsteinMulet03}, recovered here by taking $x=0$, a peak at $\eta=0$ is
observed, so that $\rho(\eta)$ can be written
$\rho(\eta)=t\delta(\eta)+(1-t)\tilde{\rho}(\eta)$ where
$\tilde{\rho}$ represents a continuous part. Rate functions
$L(\Sigma_0)$ obtained with this procedure are presented in
Fig.~\ref{fig:ldfcol}.

Interestingly, for any value of $\gamma$ clustering and SAT-UNSAT
transitions are found to occur within atypical graphs. These phase
transitions are found by monitoring the parameter $x$, which,
roughly speaking, characterizes the degree of frustration, with
larger $x$ corresponding to less frustrated graphs. For a given
$\gamma$, we indeed find thresholds $x_c(\gamma)$ and
$x_d(\gamma)$, with $x_c(\gamma)<x_d(\gamma)$, such that for
$x<x_c(\gamma)$ the graphs are UNSAT ($\Sigma_0<0$) while for
$x>x_d(\gamma)$ no more clustered solution is found
($\Sigma_0=0$). The global phase diagram in the $(\gamma,x)$ plane
is presented in Fig.~\ref{fig:phasediag}. The typical phase
diagram of Fig.~\ref{fig:cartoon} can be read on the line $x=0$,
with the thresholds $\gamma_c$ and $\gamma_d$ determined
respectively by $x_c(\gamma_c)=0$ and $x_d(\gamma_d)=0$. We also
expect that, for some values of $x$, the 1RSB Ansatz does not
hold, in analogy to what is found in the typical case
\cite{KrzakalaPagnani04} ; we however do not discuss this issue
here, which could be handled by extending the stability analysis performed for typical instances
\cite{MontanariRicci03}.

In cases such as coloring on regular graphs where the
cavity equations are factorized, an interpretation of $\F(y,\mu)$ can be given
in terms of the probability $\P_N(\e=0)$ for a graph to be SAT
while lying in a class of typically UNSAT graphs (or conversely
for a graph in a typically SAT ensemble to be UNSAT). For a RS
system first, it has been shown in \ref{sec:graphreg} that the
rate function $L(\e)$ is given by the negative 1RSB complexity
$\Sigma(\e)$, $L(\e)=-\Sigma(\e)$ ; in particular, the
probability to be SAT for a graph in the typically UNSAT phase is
$\P_N(\e=0)\asymp e^{N\Sigma_0}$, where as before
$\Sigma_0\equiv\Sigma(\e=0)$. In the typically UNSAT phase of a
1RSB system one still has $\Sigma_0<0$, but the fluctuations of
the complexity $\Sigma_0$ from graph to graph described by the
rate function $L(\Sigma_0)$ have to be taken into account. The
relation with $\P_N(\e=0)$ thus becomes
\begin{equation}
\P_N(\e=0)\asymp\int_{-\infty}^0d\Sigma_0\ e^{-N L(\Sigma_0)}
e^{N\Sigma_0}
\end{equation}
since $e^{N\Sigma_0}$ must now be multiplied by the probability
$e^{-N L(\Sigma_0)}$ to actually get a complexity $\Sigma_0$. Two
cases can then arise : the saddle point can lie at the boundary
$\Sigma_0=0$, in which case $P_N(\e=0)\asymp e^{NL(\Sigma_0=0)}$,
or it can be strictly negative, $\Sigma_0<0$, in which case
$P_N(\e=0)\asymp e^{-N[L(\Sigma_0)-\Sigma_0]}$ with $\Sigma_0$ given by
$1=\partial_{\Sigma_0}L(\Sigma_0)$. An alternative formulation can
be given with $\F(x)$ related to $L(\Sigma_0)$ through
\begin{equation}
e^{-N\F(x)}=\int d\Sigma_0\ e^{-N[L(\Sigma_0)-x\Sigma_0]}.
\end{equation}
Since $\Sigma_0=\partial_x\F(x)$, one has to compute the $x^*$
maximizing $\F(x)$, which is also associated to the saddle point
$x^*=\partial_{\Sigma_0}L(\Sigma_0=0)$ : $x^*<1$ corresponds to
the first case with $\P_N(\e=0)\asymp e^{-N\F(x^*)}$, while
$x^*>1$ corresponds to the second case with $\P_N(\e=0)\asymp
e^{-N\F(1)}$ (the very same mechanism underlies the selection of
the 1RSB parameter in the typical cavity method at finite
temperature \cite{MezardParisi01}). Physically, this second
situation, $x^*>1$, refers to very rare graphs with extremely
small frustration, which are thus in a RS phase [it can be seen
that $\F(x=1)$ indeed gives back the RS rate function
$L(\e=0)$]. In the phase diagram of a problem like coloring, such
a situation is expected only for the largest values of $\gamma$.

As such, the interpretation applies only for models where the
factorization holds. Otherwise negative complexities
$\Sigma(\e)<0$ should be interpreted as giving the probability
$e^{N\Sigma(\e)}$ for a model with typical complexity to have
a cluster with energy $\e$, which, because of the interference between the internal disorder and the local external disorder, is not associated with a
rate function relative to the external disorder only. The
estimation of $\P_N(\e=0)$ along the lines presented above is
however still amenable, but one has to perform a two-step large
deviations analysis involving a second rate function $L(L_0)$ over
the first rate function $L_0$ estimated under the RS assumption ;
technically, this computation is quite similar to what has been
done here for $L(\Sigma_0)$, which can be taken as a
factorized approximation of $L(L_0)$.

\subsection{Multiple sources of disorder}\label{sec:2step}

In problems other that
vertex-cover or coloring, the definition of an instance can
include some quenched values of the coupling constants, as happens
for spin-glass models or $K$-SAT. In this case, the energy shifts
include a dependence on the couplings $J$, with functions $\Delta
\hat{E}^{(k;\{J_i\})}_n(\{h_i\})$ and $\Delta
\hat{E}^{(J_{12})}_\ell(h_1,h_2)$. The situation is
described by a two-step large deviations principle,
\begin{equation}
\P_G[E_{G,J}=N\e]\asymp e^{N\L_G(\e)},\quad\quad \P[\L_G]\asymp e^{N\mathcal{K}(\L_G)}.
\end{equation}
The rate functions are again computed through their Legendre transforms by considering two temperatures, $y_J$ and $y_G$,
\begin{eqnarray}
\F(y_J,y_G)=& y_G f-\mathcal{K}(f,y_J),& y_G=\partial_{f}\mathcal{K}(f,y_J),\\
y_J f(y_J)=& y_J \e -\L(\e),&y_J =\partial_\e\L(\e).
\end{eqnarray}
where the factor $y_J$ in front of $f$ in the second line is introduced as in Eq.~(\ref{eq:pot2}) to conform with the traditionnal definition of free energies. Taking Poissonian graphs as an example, we have explicitely the following expressions, to be compared with Eqs.~(\ref{eq:rsb}) and (\ref{eq:rsb2}):
\begin{equation}
\begin{split}
&\mathcal{P}[P_0]=\frac{1}{\Z}\sum_{k=0}^\infty
\pi_\gamma(k)\int\prod_{i=1}^k\D P_i
\mathcal{P}[P_i]\delta\left[P_0-\hat{P}^{(k)}[\{P_i\}]\right]\hat{Z}^{(k)}[\{P_i\}]^{y_G/y_J}e^{-z\chi(k,\gamma)},\\
&P_0(h_0)=\hat{P}^{(k)}[\{P_i\}](h_0)=\frac{1}{Z}\E_J\int\prod_{i=1}^kdh_iP_i(h_i)\delta(h_0-\hat{h}^{(k,\{J_i\})}(\{h_i\}))
e^{-y_J\Delta\hat{E}_n^{(k;\{J_i\})}(\{h_i\})},\\
&Z=\hat{Z}^{(k)}[\{P_i\}]=\E_J\int\prod_{i=1}^kdh_iP_i(h_i)e^{-y_J\Delta\hat{E}_n^{(k;\{J_i\})}(\{h_i\})},\\
& e^z=\sum_{k=0}^\infty\sigma(k)\left[\int \prod_{i=1}^2\D
P_i\mP[P_i]\left(\E_J\int \prod_{i=1}^2 dh_i P_i(h_i)e^{-y_J\Delta
\hat{E}_\ell^{(J_{12})}(h_1,h_2)}\right)^{y_G/y_J}\right]^k,\\
&\F(y_J,y_G,\gamma)=-\ln\Z=-\ln\left[\sum_{k=0}^\infty
\pi_\gamma(k)\int\prod_{i=1}^k\D P_i
\mathcal{P}[P_i]\delta\left[P_0-\hat{P}^{(k)}[\{P_i\}]\right]\hat{Z}^{(k)}[\{P_i\}]^{y_G/y_J}e^{-z\chi(k,\gamma)}\right].
\end{split}
\end{equation}
where $\mathbb{E}_J$ denotes the average over the couplings $J$.
Note that with $y_G=0$, we obtain large deviations with respect to
the couplings on a typical graph, while with $y_J=0$, we obtain
large deviations with respect to the graphs for typical couplings.
With $y_G=y_J$, the two sources of
disorder are treated at a same level, in analogy with $y=\beta$ in
Sec.~\ref{sec:finiteT} and $\mu=y$, ($x=1$) in Sec.~\ref{sec:rsb} 
(this prescription is sometimes referred to as the "Nishimori temperature"~\cite{Nishimori01}).

An interesting feature of the cavity equations is the possibility
to implement them as a message-passing algorithm~\cite{MezardZecchina02} to study for example here the large deviations with respect to the couplings on a given graph. The message passed along the
oriented edge $(i\to j)$ is the distribution $P^{(i\to j)}$,
which, in particular cases, can be parameterized by a finite
set of real numbers, and the update rule is
\begin{equation}
P^{(i\to 0)}(h_i)=\frac{1}{Z}\E_J\int\prod_{j\in i-0}dh_j P^{(j\to
i)}(h_j)\delta\left(h_0-\hat{h}^{(k,\{J_{ji}\})}(\{h_j\})\right)
e^{-y_J\Delta\hat{E}_n^{(k;\{J_{ji}\})}(\{h_j\})}
\end{equation}
where the notation $j\in 0-i$ means that, on the particular graph
considered, $j$ is a neighbor of $0$ different from $i$.
This algorithmic scheme could be used to design graphs with optimal properties,
with for instance applications in coding theory, in the
context of low-density parity-check codes
\cite{RichardsonShokrollahi01}.

\section{Conclusion}

While statistical physics of disordered systems have so far mostly
focused on the thermodynamical properties of samples which are typical with
respect to the source of quenched disorder, we have shown here
that its methods can be extended to describe large
deviations, that is, atypical samples. Large deviations are of
foremost interest in probability theory and the approach followed
here, though admittedly non rigorous, is based on clearly
formulated assumptions which should be amenable to mathematical
justifications. In its simplest form indeed, the LDCM
we exposed assumes no replica symmetry breaking, a
situation in which many of its typical predictions have been
proved to be exact~\cite{Talagrand03}.

From the perspective of algorithmic complexity, the LDCM can be seen as a first
step in an attempt to reconcile the worst case and typical case
analysis, usually regarded as antagonistic. However the scope of
large deviations should not be regarded as restricted to
optimization theory, as it notably allows to work out the
statistical mechanics of physical systems with adaptive
structures. An example of such system is constituted by random
networks subject to mechanical constraints where the possibility
to adapt leads to the occurrence of an intriguing intermediate
phase, preceding a rigidity transition~\cite{BarreBishop05b} ; an
other example along the same lines is given by proteins whose structure is shaped by
strong constraints~\cite{ThorpeRader01}. Adaptative structures in random graphs are also
of interest in the seemingly unrelated field
of neural networks~\cite{WemmenhoveSkantzos04}. 
Finally, in the
spirit of the most impressive achievements of its typical
counterpart~\cite{MezardZecchina02}, it would also be interesting
to implement the LDCM on particular
ensembles of instances to analyze, and possibly design, graph
structures with specific properties.

\subsection*{Acknowledgments}

I am grateful to Andrea Montanari for providing me with his notes
\cite{Montanari02} on the analysis of large deviations by the
replica method, and to Marc M\'ezard for his interest and encouragements.

\bibliographystyle{apsrev}

\bibliography{reference,graphs,glasses}

\begin{thebibliography}{45}
\bibitem[{foo({\natexlab{a}})}]{footnoteSUSY}
\bibinfo{note}{This disability originates from the presence of marginally
  stable states and can be overpassed by considering a generalization
  (supersymmetry breaking version) of the cavity method \cite{Rizzo04}.}
\bibitem[{foo({\natexlab{b}})}]{footnoteZero}
\bibinfo{note}{Strictly speaking, a problem is SAT if and only if $E=0$, and it
  may be that $E>0$ while $\e=0$ in the thermodynamical limit, as observed for
  the 2-coloring problem \cite{MourikSaad02}. Here however, we take the
  viewpoint of defining a SAT graph by $\e=0$ : this condition in fact turns
  out to be equivalent to $E=0$ for the 3-coloring problem.}
\end{thebibliography}

\end{document}